\documentclass[]{siamltex1213}

\usepackage{color}
\usepackage{graphicx}
\usepackage{amssymb,amsmath}
\usepackage{amsfonts}
\usepackage{hyperref}

\newcommand{\sslash}{\mathbin{/\mkern-4mu/}}

\title{Vortex Nucleation in a Dissipative Variant of the Nonlinear
Schr{\"o}dinger Equation under Rotation}

\author{%
R. Carretero-Gonz{\'a}lez\footnotemark[2]
\and
P.G. Kevrekidis\footnotemark[3]
\and
T. Kolokolnikov\footnotemark[4]
}

\begin{document}
\maketitle
\slugger{mms}{xxxx}{xx}{x}{x--x}

\renewcommand{\thefootnote}{\fnsymbol{footnote}}
\footnotetext[2]{%
Nonlinear Dynamical System Group,
({\texttt{URL}: \href{http://nlds.sdsu.edu}{http:$\sslash$nlds.sdsu.edu}}),
Computational Science Research Center,
({\texttt{URL}: \href{http://www.csrc.sdsu.edu}{http:$\sslash$www.csrc.sdsu.edu}}),
and Department of Mathematics and Statistics,
San Diego State University,
San Diego, CA 92182-7720, USA}
\footnotetext[3]{%
Department of Mathematics and Statistics,
University of Massachusetts,
Amherst, MA 01003-4515, USA and
Center for Nonlinear Studies and Theoretical Division, Los Alamos
National Laboratory, Los Alamos, NM 87544}
\footnotetext[4]{%
Department of Mathematics and Statistics,
Dalhousie University Halifax,
Nova Scotia, B3H3J5, Canada}
\renewcommand{\thefootnote}{\arabic{footnote}}

\begin{abstract}
In the present work, we motivate and explore the dynamics of a dissipative
variant of the nonlinear Schr{\"o}dinger equation under the impact of
external rotation. As in the well established Hamiltonian case, the rotation
gives rise to the formation of vortices. We show, however, that the most
unstable mode leading to this instability scales with an appropriate
power of the chemical potential $\mu$ of the system, increasing
proportionally to $\mu^{2/3}$. The precise form of the relevant formula,
obtained through our asymptotic analysis, provides the most unstable mode as
a function of the atomic density and the trap strength.
We show how these unstable modes typically nucleate a large number
of vortices in the periphery of the atomic cloud.
However, through a pattern selection mechanism, prompted by symmetry-breaking,
only few isolated vortices are pulled in sequentially from the periphery 
towards the bulk of the cloud resulting in highly symmetric stable vortex 
configurations with far fewer vortices than the original unstable mode.
These results may be of relevance to the experimentally tractable realm 
of finite temperature atomic condensates.
\end{abstract}

\begin{keywords}
Vortex nucleation,
nonlinear Schr\"odinger equation,
Gross-Pitaevskii equation,
Bose-Einstein condensates.
\end{keywords}

\begin{AMS}
34A34, 
35Q55, 
76M23, 
\end{AMS}

\pagestyle{myheadings}
\thispagestyle{plain}
\markboth{R. Carretero-Gonz{\'a}lez, P.G. Kevrekidis, and T. Kolokolnikov}%
{Vortex Nucleation in a Dissipative Variant of the NLS Equation under Rotation}


\section{Introduction}

Vortices are persistent circulating flow patterns that occur in many diverse
scientific and mathematical contexts~\cite{Pismen1999}, ranging from
hydrodynamics, superfluids, and nonlinear
optics~\cite{Kivshar-LutherDavies,YSKPiO}
to specific case examples in sunspots, dust
devils~\cite{Lugt1983}, and plant propulsion~\cite{Whitaker2010}. The realm
of atomic Bose-Einstein condensates
(BECs)~\cite{review_dalfovo,becbook1,becbook2} has produced a novel and pristine setting
where numerous features of the exciting nonlinear dynamics of single- and
multi-charge vortices, as well as of vortex crystals and vortex lattices,
can be not only theoretically studied, but also experimentally observed.

The first experimental observation of vortices in atomic
BECs~\cite{Matthews99} by means of a phase-imprinting
method between two hyperfine
spin states of a $^{87}$Rb BEC~\cite{Williams99} paved the way for a
systematic investigation of their dynamical properties.
Stirring the BECs~\cite{Madison00} above a certain critical angular
speed~\cite{Recati01,Sinha01,Madison01} led to the production of few
vortices~\cite{Madison01} and even of robust vortex
lattices~\cite{Raman,jamil}. Other
vortex-generation techniques were also used in experiments, including the
breakup of the BEC superfluidity by dragging obstacles through the
condensate~\cite{kett99}, as well as nonlinear interference between
condensate fragments \cite{BPAPRL}. In addition, apart from unit-charged
vortices, higher-charged vortex structures were produced \cite{S2Ket} and
their dynamical (in)stability was examined. To these earlier experimental
developments, one can add in recent years: the formation of vortices through
a quench of a gas of atoms from well above to well-below the BEC transition
via the so-called Kibble-Zurek mechanism~\cite{BPA}; the dynamical
visualization of such \textquotedblleft nucleated\textquotedblright\
vortices~\cite{dshall} and even of vortex pairs, i.e., dipoles consisting of
two oppositely charged vortices; the nucleation of dipoles via the dragging
of a laser beam through the BEC~\cite{BPA3}; the systematic experimental
exploration of dipole dynamics~\cite{dshall1}; the generation (via
instabilities) of 3-vortex configurations of same or opposite signs
in Ref.~\cite{tripole}
and the ``dialing in'' of
arbitrary numbers of (few) same-charge vortices and the visualization of
their intriguing, potentially symmetry-breaking dynamics~\cite{dshall3}.
Naturally, the above developments suggest that the study of vortices and of
their nucleation in BECs is a theme of broad and intense ongoing
interest.

On the other hand, another topic receiving increasing attention has
concerned the role of finite-temperature induced ``damping'' of the
BEC~\cite{Proukakis_Book}.
A wide range of recent examples has indicated that this
leads to anti-damping motion of the coherent structures such as solitary
waves (dark solitons) and vortices. Early soliton experiments of about 15 years ago observed the motion of a dark soliton towards the edge of the
trap~\cite{han1,han2,nist}.
It is interesting to note, however, that this type of
anti-damping effect has been observed in a far more pronounced way in recent
experiments of dark soliton oscillations in a unitary Fermi
gas~\cite{Zwierlein}. A number of theoretical studies have provided relevant
explanation for this phenomenology in atomic
BECs~\cite{shl1,shl2,us,ft1,ft2,ft3,ft4,gk,ashton}. In particular, it has been
identified in these works that the dark soliton follows an anti-damped
harmonic oscillator behavior, leading to trajectories of growing amplitude
around the center of the trap, until expelled from the BEC. This motion has
been observed in the context of the so-called dissipative Gross-Pitaevskii
equation model (DGPE). The DGPE was originally introduced phenomenologically
by Pitaevskii~\cite{lp} as a way to use a damping term to account for the
role of finite temperature induced fluctuations in the BEC dynamics; see,
e.g., Refs.~\cite{penckw,npprev,blakie,ZNG_Book} for discussions and
microscopic interpretations of such a term. Comparisons~\cite{us} of its
results with more elaborate models such as the (averaged quantities of) the
stochastic Gross-Pitaevskii equation (SGPE)~\cite{stoof_sgpe,stoch} offered
good reason for exploring the simpler DGPE model, as regards coherent
structure (such as soliton) dynamics.

More importantly for our theme of vortex dynamics, an increasing volume of
literature has been exploring the role of thermal
effects~\cite{proukv1,proukv2,ourjpb,ashtone,tmwright}.
Here, too, and in accordance with
theoretical predictions~\cite{stoof} (see also Ref.~\cite{ashtone2} and for a
recent discussion~\cite{dongyan}), an outward, in this case spiraling,
trajectory is found for the single vortex motion which leads to its
expulsion from the trap. While important aspects of the vortex dynamics in
the presence of the thermal component such as the single
vortex motion~\cite{stoof,ashtone2} and even the vortex-pair
interaction~\cite{dongyan} have
been explored theoretically (and numerically), to the best of our knowledge,
the predictions of the DGPE model in the context of vortex nucleation under
rotation have not been previously examined.

The principal scope of our study will, thus, be to provide some insight on
the instability and dynamics that leads to the emergence of vortices in the
presence of ``thermal dissipation'', i.e., in the DGPE framework. In fact, to
facilitate the analysis, we will go one step further in simplifying the
problem and will also explore the ``imaginary time'' analogue of the GPE.
This choice will be suitably explained and motivated in the next section.
Subsequently, in Section~\ref{sec:overdampedNLS}, we will provide the analysis that identifies the
most unstable eigenmode and its scaling with the system parameters (most
notably, the chemical potential $\mu$). Finally, in Section~\ref{sec:conclu}, we will
summarize our findings and present a number of directions for future study.

\section{Model Setup: the NLS Equation Under Rotation
and its Dissipative Variant}

\subsection{Dissipationless case}

The standard GPE model valid at $T=0$ for describing the the
quasi-2D condensate wavefunction $u(x,y,t)$ in the presence of rotation is:
\begin{eqnarray}
i u_t=-\frac{1}{2} \Delta u + \frac{1}{2} \Omega_{\rm trap}^2 r^2 u -
\mu u + |u|^2 u + i \Omega_{\rm rot} u_{\theta},
\label{tk_eqn1}
\end{eqnarray}
where $(\cdot)_t=d(\cdot)/dt$ and $(\cdot)_\theta=d(\cdot)/d\theta$ and
$(r,\theta)$ are the polar coordinates.
Here the potential is assumed as representing a parabolic (typically induced
magnetically) trap of strength $\Omega_{\rm trap}$, while an external
rotation of strength $\Omega_{\rm rot}$ is assumed to be imposed.
We have also explicitly included the chemical potential $\mu$ in
the model (although it can be factored out by a gauge transformation), as it
will be relevant in the DGPE variant of the system. Notice that here we use
the dimensionless form of the pancake-shaped, 2D BEC model that has been
well established in a variety of archival references in the
field~\cite{becbook1,becbook2,emergent}.

Before we move to the DGPE variant of the model, we should note a remarkable
implication of Eq.~(\ref{tk_eqn1}). In particular, when analyzing the
spectrum of a particular state $u_0$, by performing Bogolyubov-de
Gennes (BdG) analysis to explore its stability, the equation obtained for
$u=u_0(x,y) + \epsilon v(x,y,t)$ is of the form:
\begin{eqnarray}
i v_t=-\frac{1}{2} \Delta v + \frac{1}{2} \Omega_{\rm trap}^2 r^2 v -
\mu v + 2 |u_0|^2 v + u_0^2 v^{\star} + i \Omega_{\rm rot} v_{\theta},
\label{tk_eqn2}
\end{eqnarray}
where $(\cdot)^{\star}$ denotes complex conjugation. Now, decomposing the
perturbation as 
$v(x,y,t)=a(r) e^{i m \theta} e^{i \omega t} 
        + b(r) e^{-i m \theta} e^{-i \omega t}$,
it is straightforward to see that for
a radial state (such as the ground state of the system), the sole influence
of the rotation frequency $\Omega_{\rm rot}$ is to shift the
frequencies $\omega \rightarrow \omega \pm m \Omega_{\rm rot}$.

\begin{figure}[tb]
\begin{center}
\begin{tabular}{cc}
\includegraphics[width=0.4\textwidth]{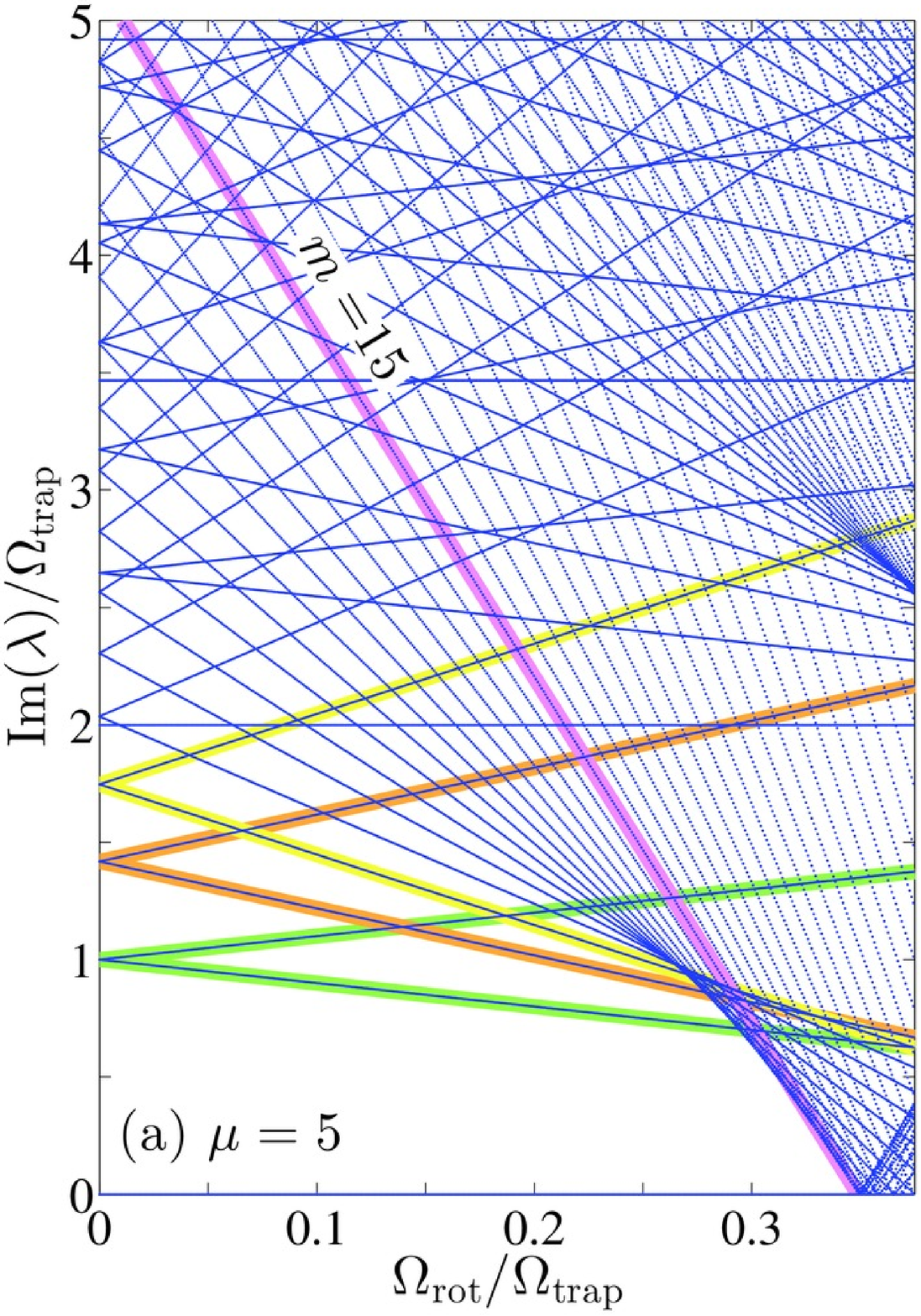}&~
\includegraphics[width=0.4\textwidth]{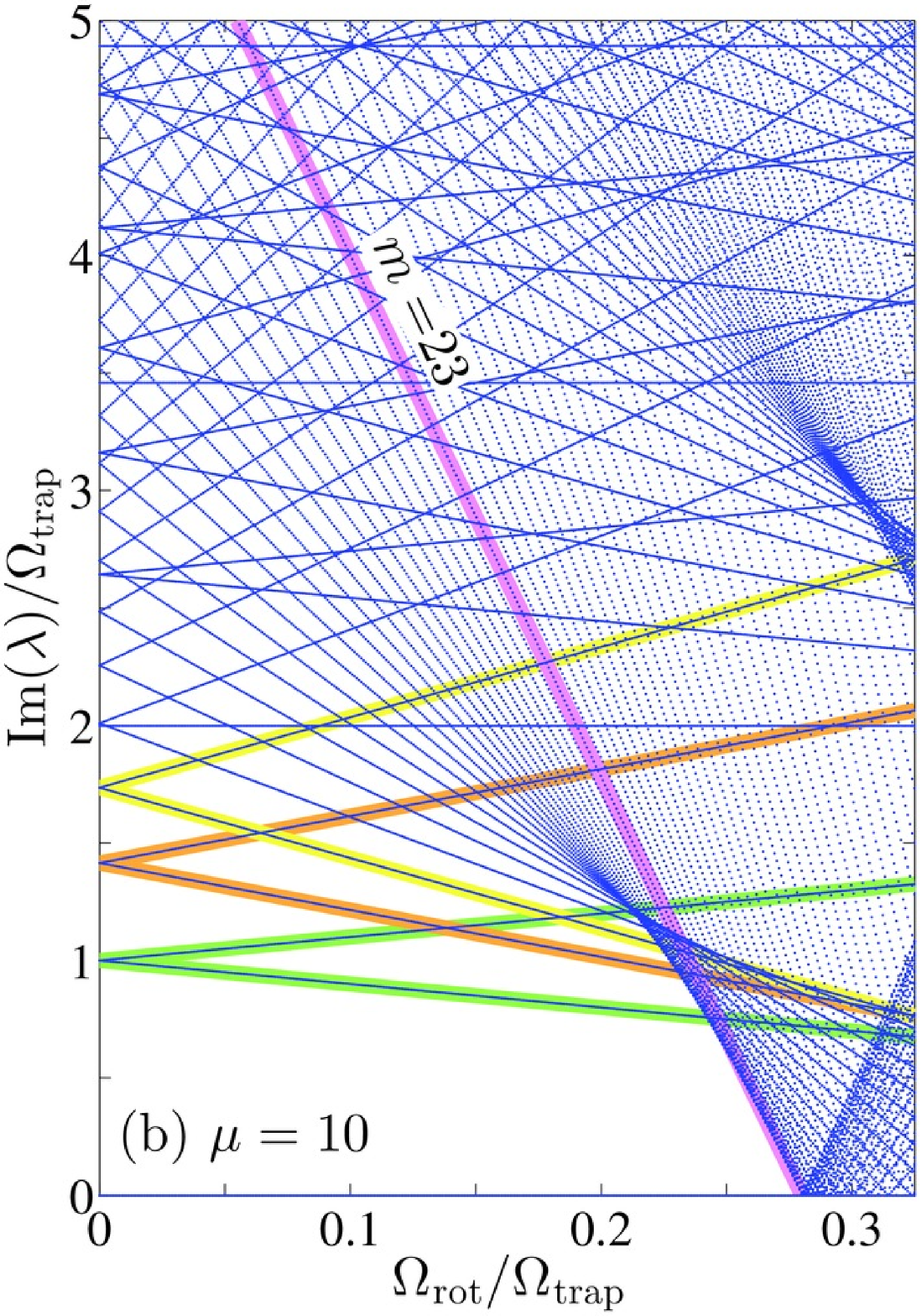}
\end{tabular}
\end{center}
\caption{(color online)
The spectrum of imaginary eigenvalues (normalized to
the trap frequency) of the Hamiltonian linearization problem
of Eq.~(\ref{tk_eqn2}) in the presence of rotation.
The left panel is for $\mu=5$, while the right one
is for $\mu=10$; $\Omega_{\rm trap}=0.2$ was chosen.
The thick green, orange and yellow lines correspond to the lowest three
modes of $m=1,2,3$, giving the theoretical prediction of
how they depend as a function of frequency according to
$\lambda= i (\omega(m=0) \pm m \Omega_{\rm rot})$.
The thick pink line corresponds to the $m=m_c$ mode that
first becomes unstable as the rotation is increased.
$m_c=15$ for $\mu=5$ and $m_c=23$ for $\mu=10$.
%
}
\label{tkfig2}
\end{figure}

This is illustrated, e.g., in Fig.~\ref{tkfig2} by direct numerical
computations involving the BdG linearization around the ground state for the
case of $\Omega_{\rm trap}=0.2$ for 2 different values of $\mu=5$
(left) and $\mu=10$ (right). The lowest, well-known modes of the condensate
dynamics namely the dipolar, quadrupolar, and hexapolar modes at,
respectively,
$\omega=\Omega_{\rm trap}$,
$\omega=\sqrt{2}\Omega_{\rm trap}$, and
$\omega=\sqrt{3}\Omega_{\rm trap}$
(all with double degeneracy), are shown by thick green, orange and yellow
lines respectively, showcasing the validity of the above eigenfrequency
shift statement. However, there are numerous additional intriguing features
to observe in the figure. For one thing, we note that since the ground state
is stable and its imaginary eigenvalues (real eigenfrequencies) shift along
the imaginary axis of the complex spectral plane (Re($\lambda$),Im($\lambda$)),
the ground state will {\em never} become dynamically unstable. Instead,
what happens is that it becomes {\em energetically} unstable acquiring
what is known as negative energy modes~\cite{skryabin} or in the
mathematical literature as negative Krein signature modes~\cite{sandstede}.
These modes indicate that while the solution may not be dynamically
unstable, it is no longer the ground state of the system. Moreover, if a
pathway, such as the presence of dissipation, becomes 
available for {\em relaxing}
to the ground state of the system then it would do so~\cite{wuniu}.

Some additional observations are also in order. In particular, it is
worthwhile to note that among the modes crossing zero to become negative
energy or signature ones, it is {\em neither} the $m=1$, {\em nor} the
$m=2$ ones that do this the first. Instead, modes associated with higher $m$
(but which start at larger $\omega$) values move faster and cross $0$,
spearheading the energetic instability of the present state.
For instance, as is depicted in Fig.~\ref{tkfig2}, the modes
with $m=m_c=15$ and $m=m_c=23$ are the first modes to
cross the energetic stability threshold for, respectively,
$\mu=5$ and $\mu=10$. This instability occurs at
$\Omega_{\rm rot}=\Omega_{{\rm rot},c} \approx 0.349\, \Omega_{\rm trap}$
for $\mu=10$ and at
$\Omega_{\rm rot}=\Omega_{{\rm rot},c} \approx 0.28\, \Omega_{\rm trap}$
for $\mu=5$.
Notice that the critical rotation threshold for $\mu=5$ is larger
that the one for $\mu=10$,
an important feature to which we will return below. The
reason why the above observations are especially interesting is the
following. As proved rigorously in Ref.~\cite{sandstede}, the inclusion of
dissipation in a Hamiltonian model leads modes of different energy (Krein
signature) to move differently, due to their distinct topological
characteristics. More specifically, modes with positive signature move to
the left of the spectral plane becoming stable/attracting eigendirections
for the dynamics. However, modes with negative Krein signature move
{\em in the opposite direction} of the spectral plane, namely to the right hand
plane, becoming immediately unstable as soon as the dissipation is turned
on. This statement goes hand-in-hand with the opening of relaxation
channels through which the solution can now revert to its preferred ground
state equilibrium, given its energetic instability.

\begin{figure}[tb]
\begin{center}
\begin{tabular}{cc}
\includegraphics[height=5.0cm]{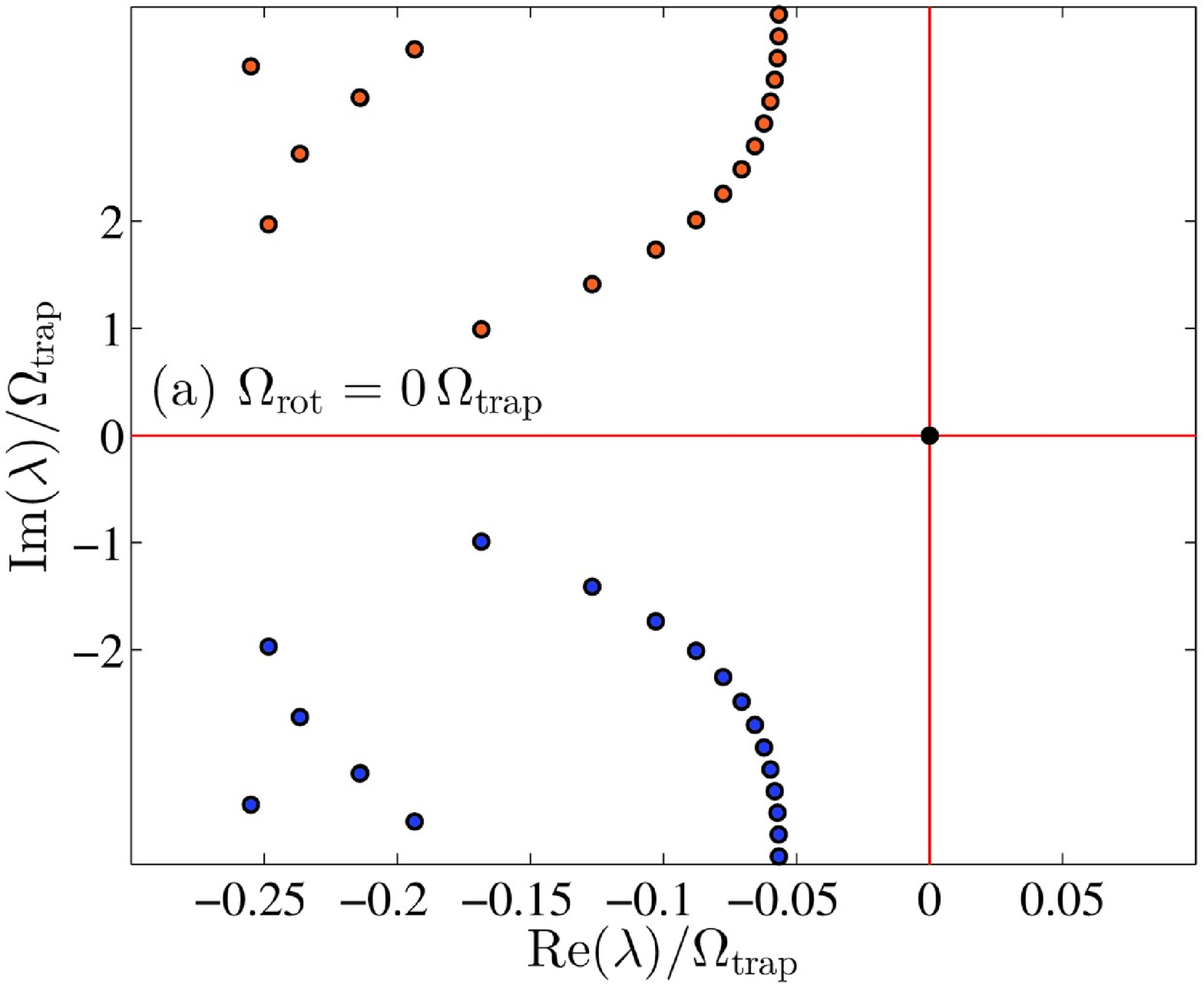}&
\includegraphics[height=5.0cm]{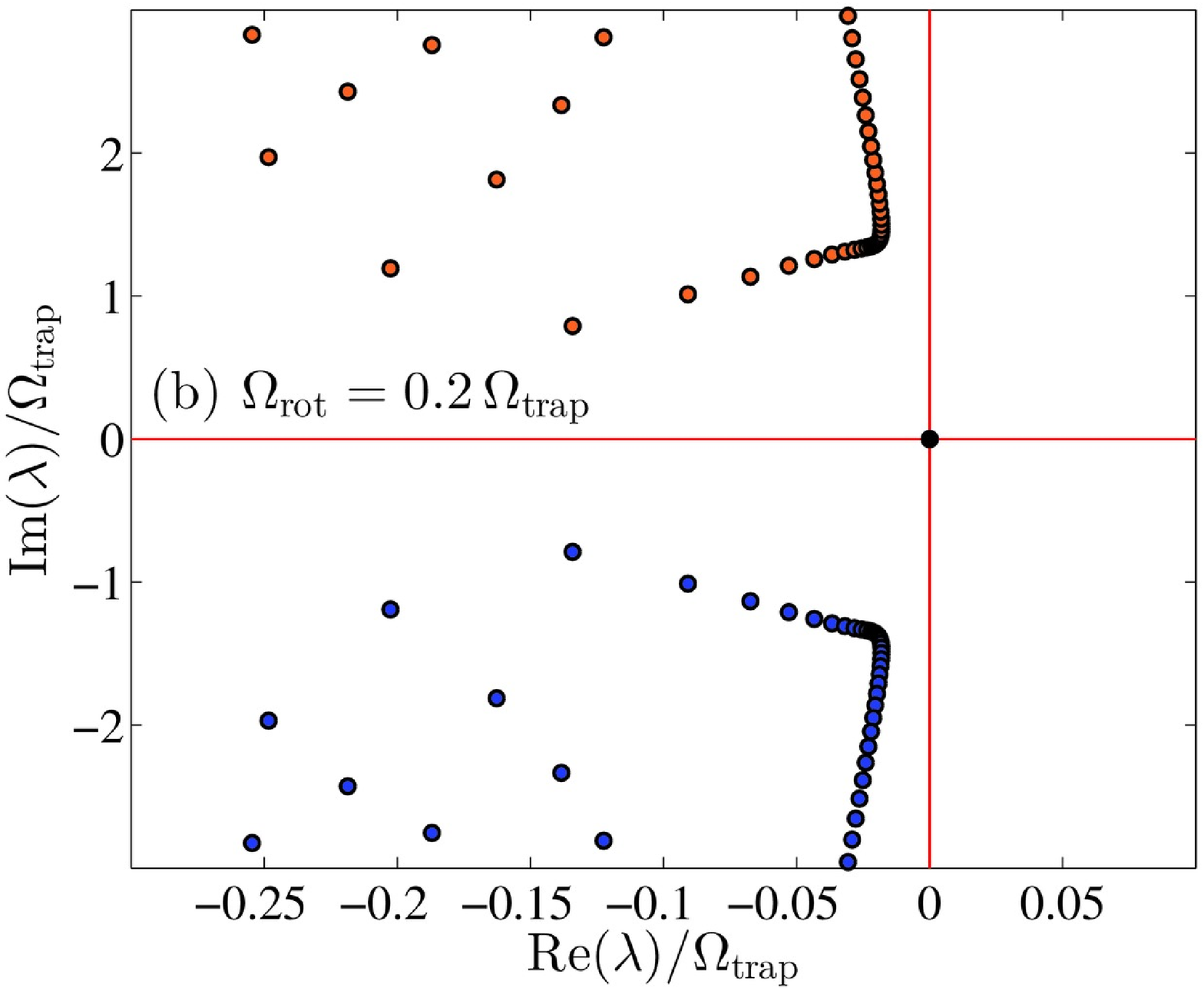}\\[0.5ex]
\includegraphics[height=5.0cm]{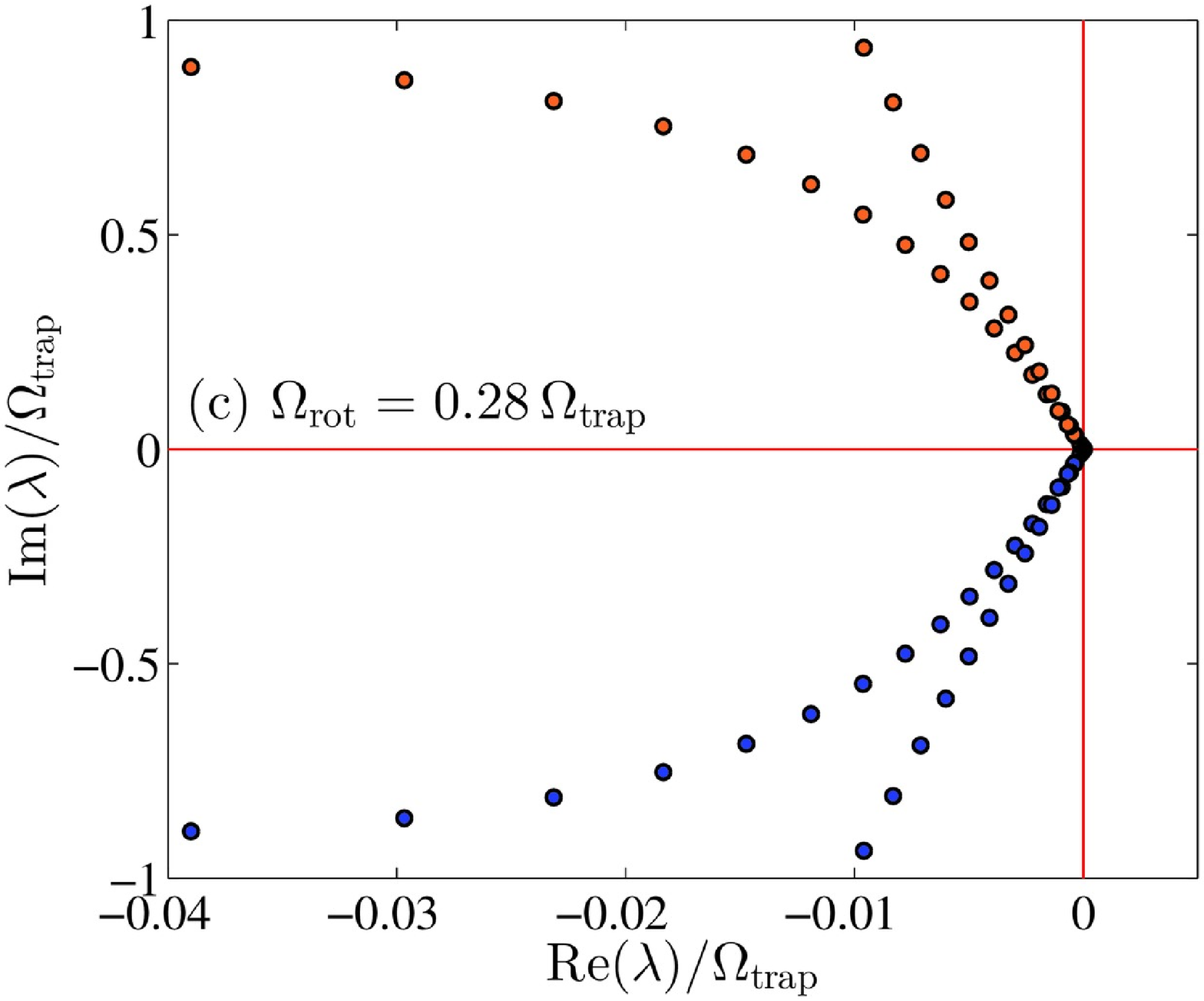}&
\includegraphics[height=5.0cm]{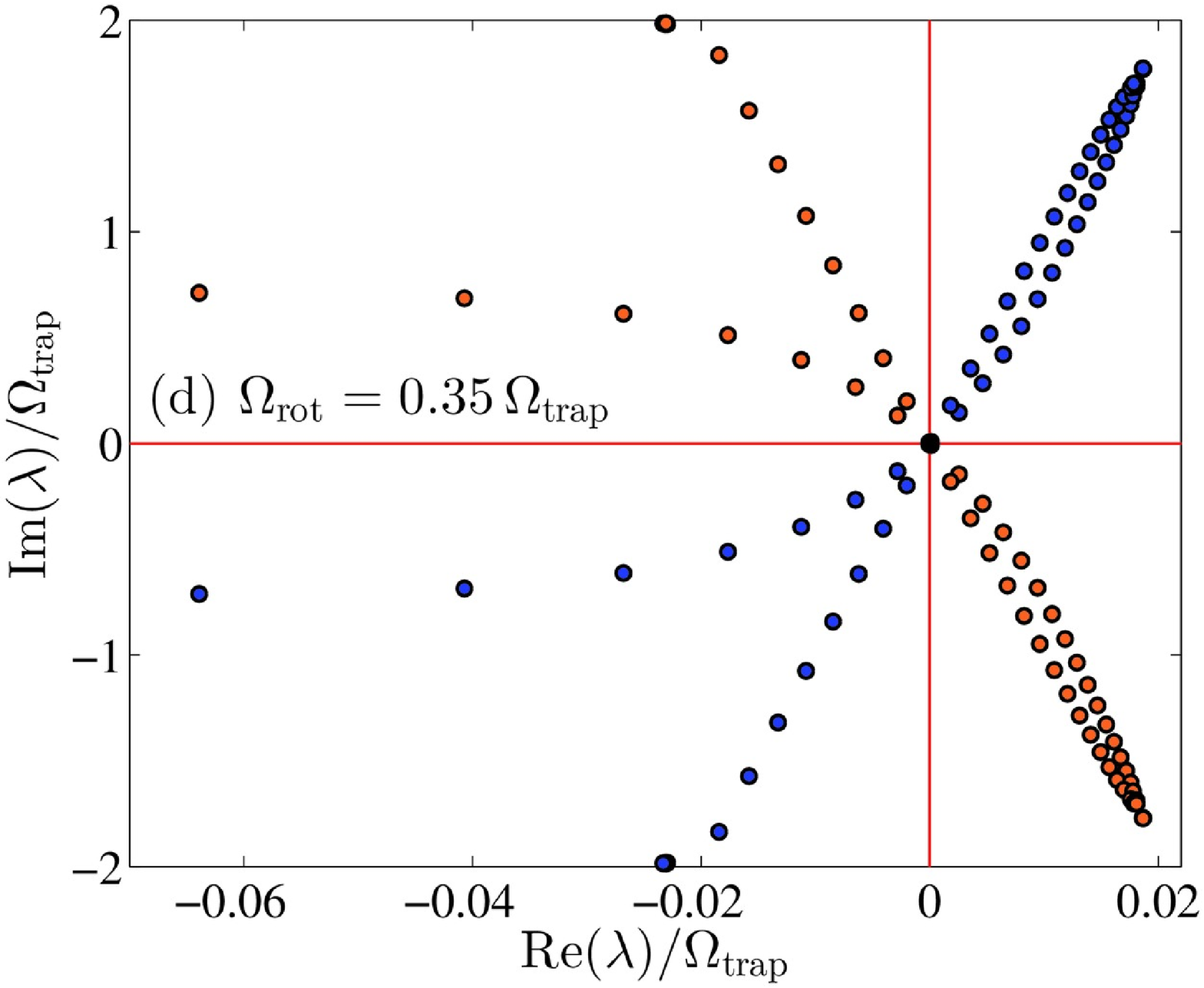}
\end{tabular}
\end{center}
\caption{(color online)
Spectral planes increasing rotation:
(a) $\Omega_{\rm rot}=0$,
(b) $\Omega_{\rm rot}/\Omega_{\rm trap}=0.2$,
(c) $\Omega_{\rm rot}/\Omega_{\rm trap}=0.28$ and
(d) $\Omega_{\rm rot}/\Omega_{\rm trap}=0.35$,
all for $\gamma=0.01$ and $\mu=10$.
Eigenvalues with positive and negative Krein sign are depicted,
respectively, with blue (dark) and orange (light) points and
zero-Krein eigenvalues are depicted with black points.
The successive panels clearly illustrate the instability due to
the collision of opposite Krein signature modes that starts
at $\Omega_{{\rm rot},c}= 0.28\, \Omega_{\rm trap}$ [see panel (c)].
}
\label{tkfig3}
\end{figure}

\subsection{Dissipative Gross-Pitaevskii Equation}

\begin{figure}[tb]
\begin{center}
\includegraphics[width=9cm]{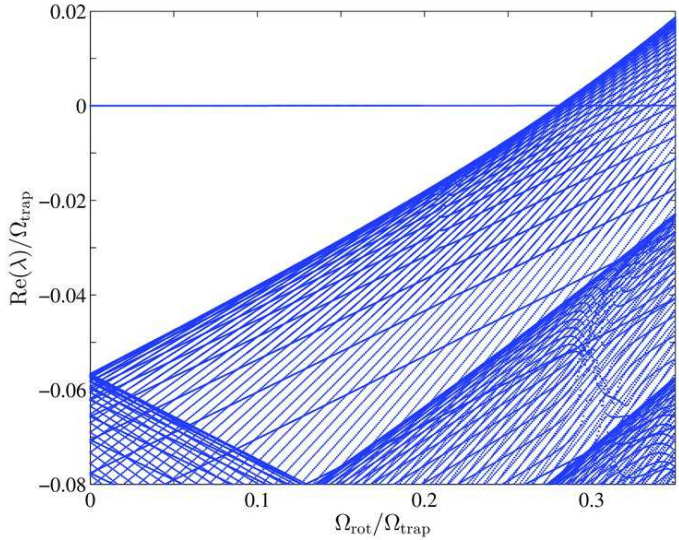}
\end{center}
\caption{(color online)
The largest eigenvalues leading to the instability beyond the critical
value of $\Omega_{\rm rot}/\Omega_{\rm trap}$, in the DGPE case
with $\gamma=0.01$.}
\label{tkfig4}
\end{figure}

\begin{figure}[t]
\begin{center}
\includegraphics[width=6.25cm]{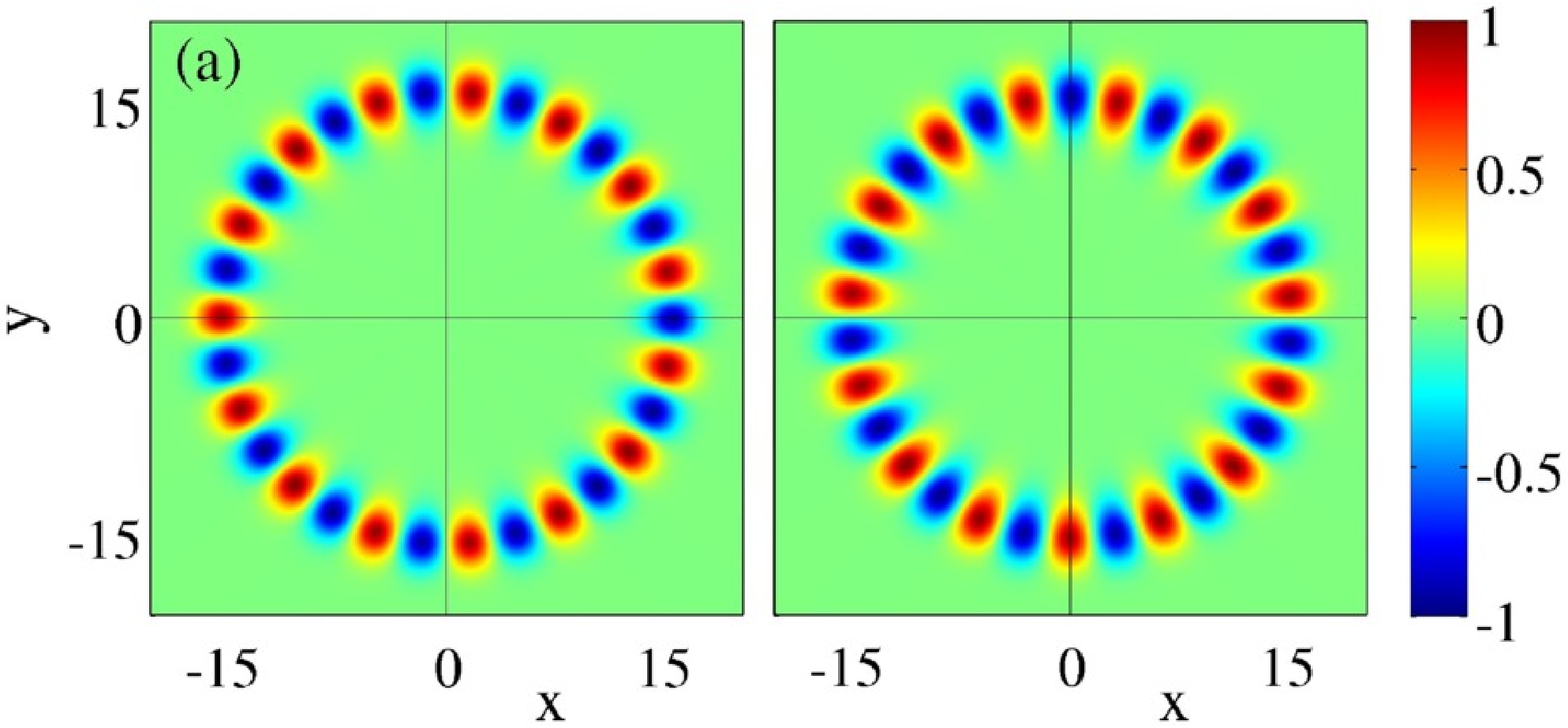}
~
\includegraphics[width=6.25cm]{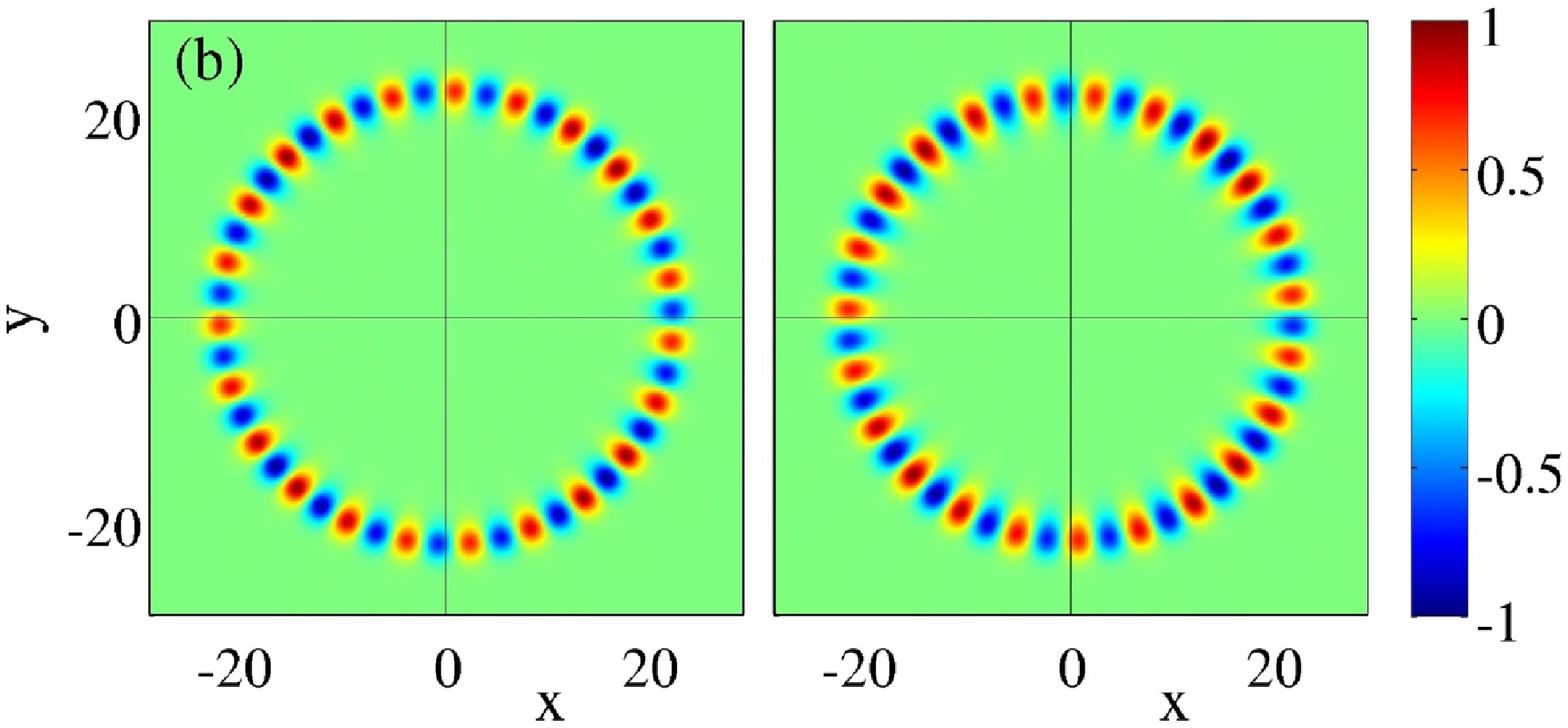}
\end{center}
\caption{(color online)
The most unstable eigenfunction just past the rotation threshold.
The left and right subpanels corresponds, respectively, to the
real and imaginary parts of the most unstable eigenfunction.
(a) For $\mu= 5$ and $\Omega_{\rm rot}/\Omega_{\rm trap}=0.3495 >
\Omega_{{\rm rot},c}/\Omega_{\rm trap}\approx 0.349$
the most unstable eigenfunction corresponds to $m=15$.
(b) For $\mu=10$ and $\Omega_{\rm rot}/\Omega_{\rm trap}=0.2802 >
\Omega_{{\rm rot},c}/\Omega_{\rm trap}\approx 0.28$
the most unstable eigenfunction corresponds to $m=23$.
}
\label{fig:eigenfun}
\end{figure}

Now, let us project the above conclusions to the case of the DGPE which is
of the form~\cite{lp}:
\begin{eqnarray}
(i-\gamma) u_t=-\frac{1}{2} \Delta u + \frac{1}{2} \Omega_{\rm trap}^2
r^2 u - \mu u + |u|^2 u + i \Omega_{\rm rot} u_{\theta},
\label{eq:DGPE}
\end{eqnarray}
where $\gamma$ ($>0$) refers to the temperature dependent parameter that has been
discussed extensively in this
context~\cite{Proukakis_Book,penckw,npprev,blakie,ZNG_Book}.
Exploring a nearly realistic
(although slightly higher than relevant, for illustration purposes; see, e.g.,
the discussion of Ref.~\cite{dongyan}) value of $\gamma=0.01$, we obtain the
results for $\mu=10$ illustrated in Figs.~\ref{tkfig3} and \ref{tkfig4}. The
former one shows the spectral planes for 4 values of the ratio
$\Omega_{\rm rot}/\Omega_{\rm trap}$,
two below, one (approximately) at, and one
above the energetic instability threshold of the ground state.
In the presence of the $\gamma$
term (irrespectively of however small), when all the modes are positive
energy ones, i.e., below the threshold of
$\Omega_{{\rm rot},c}= 0.28\, \Omega_{\rm trap}$, all of the eigenvalues
are on the left half plane [see Figs.~\ref{tkfig3}(a) and (b)], hence the
configuration is dynamically stable as well (in the DGPE case).
{\em However}, above the energetic instability threshold for the Hamiltonian
problem, the existence of negative Krein signature modes immediately leads
to the bifurcation of unstable eigenmodes in the right half of the spectral
plane [see Fig.~\ref{tkfig3}(d)] and the configuration is
dynamically unstable for the DGPE. This instability is manifested as a
function of $\Omega_{\rm rot}/\Omega_{\rm trap}$ for the DGPE case
in Fig.~\ref{tkfig4} showcasing that the nontrivial real parts of the
relevant eigenvalues emerge as the threshold is crossed.
To complement the instability picture, we depict in
Fig.~\ref{fig:eigenfun} the most unstable modes for rotations
just above the instability threshold. For $\mu=5$
[see Fig.~\ref{fig:eigenfun}(a)] the most unstable
eigenfunction for $\Omega_{\rm rot}/\Omega_{\rm trap}=0.3495$ (i.e., just
above the threshold $\Omega_{{\rm rot},c}/\Omega_{\rm trap}\approx 0.349$)
is the mode with $m=15$ as expected from Fig.~\ref{tkfig2}.
Similarly, for $\mu=10$ [see Fig.~\ref{fig:eigenfun}(b)], the most unstable
eigenfunction for $\Omega_{\rm rot}/\Omega_{\rm trap}=0.2802$ (i.e., just
above the threshold $\Omega_{{\rm rot},c}/\Omega_{\rm trap}\approx 0.28$)
is the mode with $m=23$ as expected from Fig.~\ref{tkfig2}.

\begin{figure}[ht]
\begin{center}
\includegraphics[width=11cm]{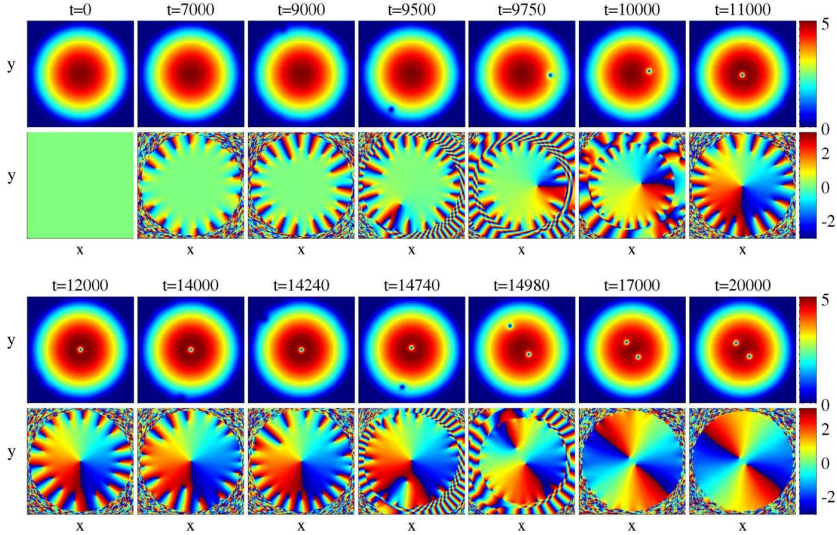}
\end{center}
\caption{(color online)
Evolution of an initial steady state without vortices
under rotation. The parameters are
$\mu=5$, $\gamma=0.01$, $\Omega_{\rm trap}=0.2$, and $\Omega_{\rm rot}/\Omega=0.37$.
For each of the times indicated, we depict the density (top sub-rows) and
phase (bottom sub-rows). The windows for density and phase are,
respectively, $(x,y)\in[-16,16]\times[-16,16]$ and
$(x,y)\in[-29.5,29.5]\times[-29.5,29.5]$.
We invite the interested reader to see the full movie at this address:
\href{http://nonlinear.sdsu.edu/~carreter/RotatingBEC.html}%
{http:$\sslash$nonlinear.sdsu.edu/~carreter/RotatingBEC.html} [Movie\#1].
}
\label{fig:snap_film_mu05}
\end{figure}

\begin{figure}[ht]
\begin{center}
\includegraphics[width=11cm]{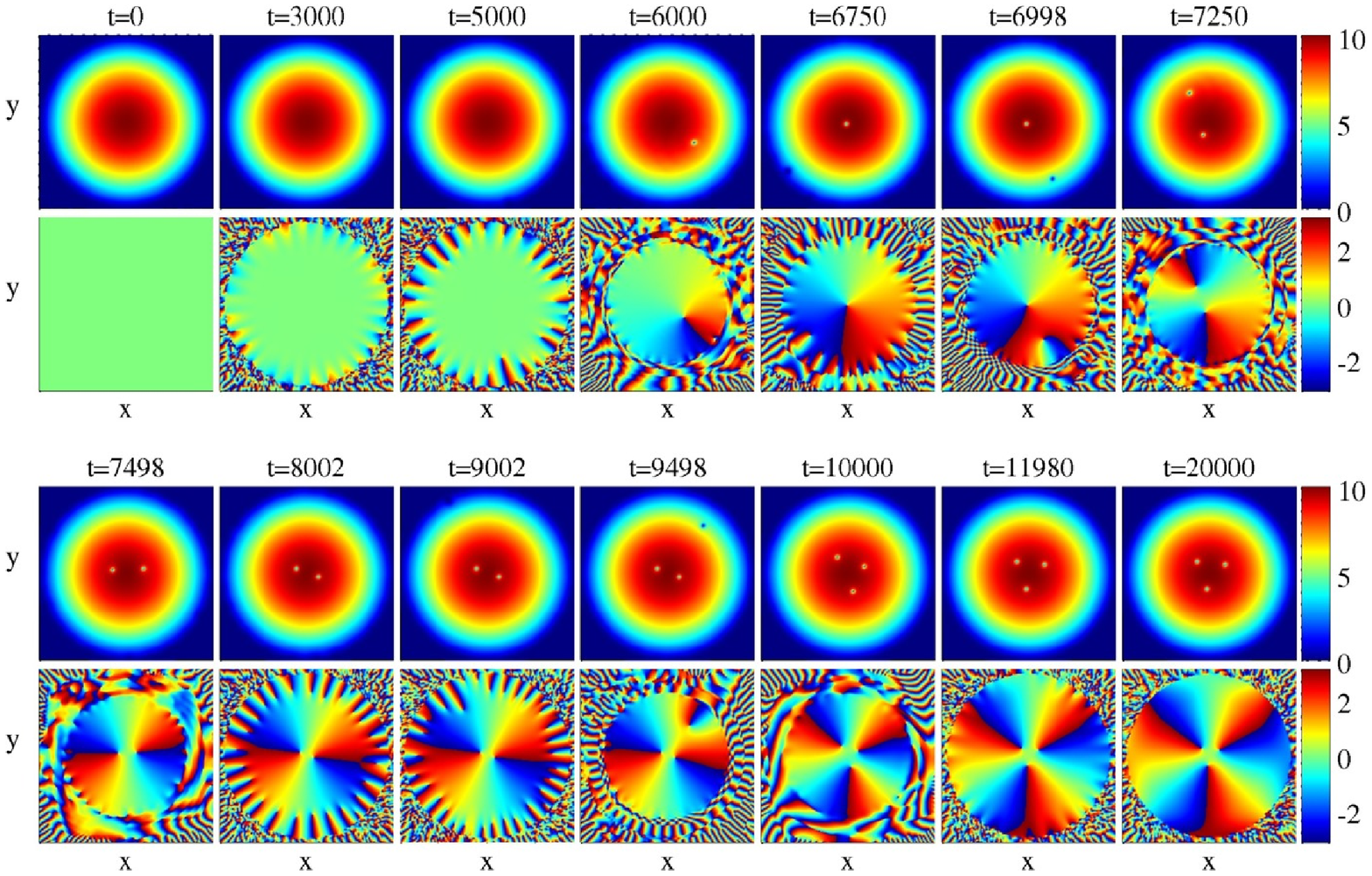}
\end{center}
\caption{(color online)
Same as in Fig.~\ref{fig:snap_film_mu05} for
$\Omega_{\rm rot}/\Omega_{\rm trap}=0.3$ and $\mu=10$.
The windows for density and phase are,
respectively,  $(x,y)\in[-22.5,22.5]\times[-22.5,22.5]$ and
$(x,y)\in[-35.8,35.8]\times[-35.8,35.8]$.
We invite the interested reader to see the full movie at this address:
\href{http://nonlinear.sdsu.edu/~carreter/RotatingBEC.html}{http:$\sslash$nonlinear.sdsu.edu/~carreter/RotatingBEC.html} [Movie\#2].
}
\label{fig:snap_film_mu10}
\end{figure}

The above stability considerations allow us to understand the
bifurcations (instabilities) of steady states bearing no initial
vorticity as the rotation of the BEC cloud is increased.
In particular, a number $N_v$ of vortices will be nucleated at the
periphery of the cloud through an unstable eigenfunction 
invariant under rotations by $2\pi/N_v$.
See for example the eigenfunctions depicted in Fig.~\ref{fig:eigenfun}.
However, it is crucial to note that 
this analysis only captures the initial stages of the dynamical evolution
and the eventual asymptotic behavior may well be different.
This can be due to symmetry-breaking effects generated by
infinitesimally small, non-symmetric, perturbations that will
generically be present in physical (and numerical) setups.
Therefore, we now explore the dynamics of the above mentioned
unstable modes towards understanding what they actually nucleate
as the instability sets in.
For this purpose we have produced long-term simulations of
the DGPE~(\ref{eq:DGPE}) starting from the stationary state
bearing no vorticity.
Two typical evolutions are depicted in Figs.~\ref{fig:snap_film_mu05}
and \ref{fig:snap_film_mu10} for, respectively, $\mu=5$ and $\mu=10$.
We invite the interested reader to see the full movies at this address:
\href{http://nonlinear.sdsu.edu/~carreter/RotatingBEC.html}{http:$\sslash$nonlinear.sdsu.edu/~carreter/RotatingBEC.html} [Movies\#1 and \#2].
The simulations were chosen for rotations that are slightly above
critical so that the steady state with no vortices is (weakly) unstable.
Let us describe in detail the full evolution for the first case
with $\mu=5$ for $\Omega_{\rm rot}=0.37\, \Omega_{\rm trap} >
\Omega_{{\rm rot},c} \approx 0.349\, \Omega_{\rm trap}$ that is
depicted in Fig.~\ref{fig:snap_film_mu05}.
As it is clear from the figure, for the chosen parameter values,
the steady state with no vortices is unstable towards a mode with
$m=17$ vortices initially growing at the periphery of the cloud
(see snapshot at $t=7,000$).
This mode is not apparent in the density distribution since it is
outside the Thomas-Fermi radius where the density is too low to
be able to be picked up. However, the phase distribution clearly
shows a series of $2\pi$ windings that are nucleated at the periphery.
It is clear that the growth of this unstable mode is prone to
asymmetries since it is generated numerically from the 
noise inherent in the computation due to its finite precision. This effect would be similar
in the physical experiments where small variations in the initial density
and the trapping break the symmetry of the solution.
This asymmetry is responsible for one of the vortices to be closer
to the center of the cloud than its siblings. This selection
mechanism is responsible for one of the vortices to start spiralling
inwards (see snapshots $t=9,000$--$11,000$). It is interesting that,
as the chosen vortex rotates close to its siblings, it ``pushes'' the
other vortices outwards and thus further contributes to this
selection mechanism.
After the chosen vortex relaxes at the center of the trap, another
unstable mode at the periphery grows and, by the same selection
mechanism explained above, spirals inwards
(see snapshots at $t=12,000$--$17,980$).
Then, the two central vortices arrange themselves into a
steady state configuration (see snapshot at $t=17,000$)
with a small perturbation that remains at the periphery.
However, in this case, this state is no longer unstable
and hence the dynamics is eventually attracted to it 
(see snapshots at $t=17,000$--$20,000$).
In fact, the resulting state with two corotating vortices in
completely stable and thus the configuration relaxes towards it
and remains there.

A similar evolution is observed in the case of $\mu=10$ and
$\Omega_{\rm rot}=0.3\, \Omega_{\rm trap} >
\Omega_{{\rm rot},c} \approx 0.28\, \Omega_{\rm trap}$ that is
depicted in Fig.~\ref{fig:snap_film_mu10}.
In this case, the state with two vortices in the bulk of the
condensate is still unstable and thus a third vortex needs
to be pulled from the periphery inwards to finally create a
corotating tripole that is spectrally {\em stable}.

\begin{figure}[tb]
\begin{center}
\begin{tabular}{ccc}
\hskip-0.1cm
\includegraphics[height=3.56cm]{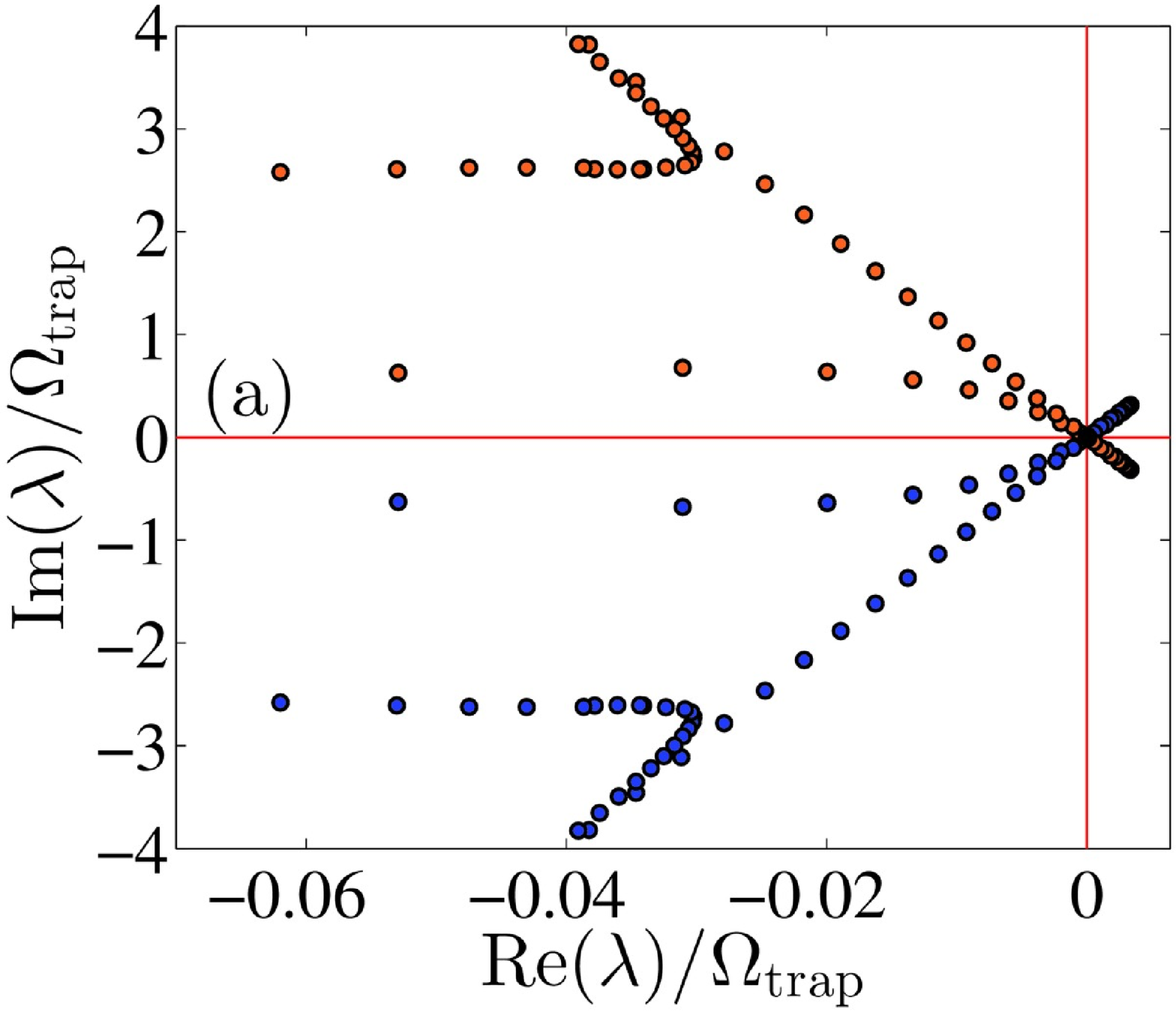}&
\hskip-0.3cm
\includegraphics[height=3.56cm]{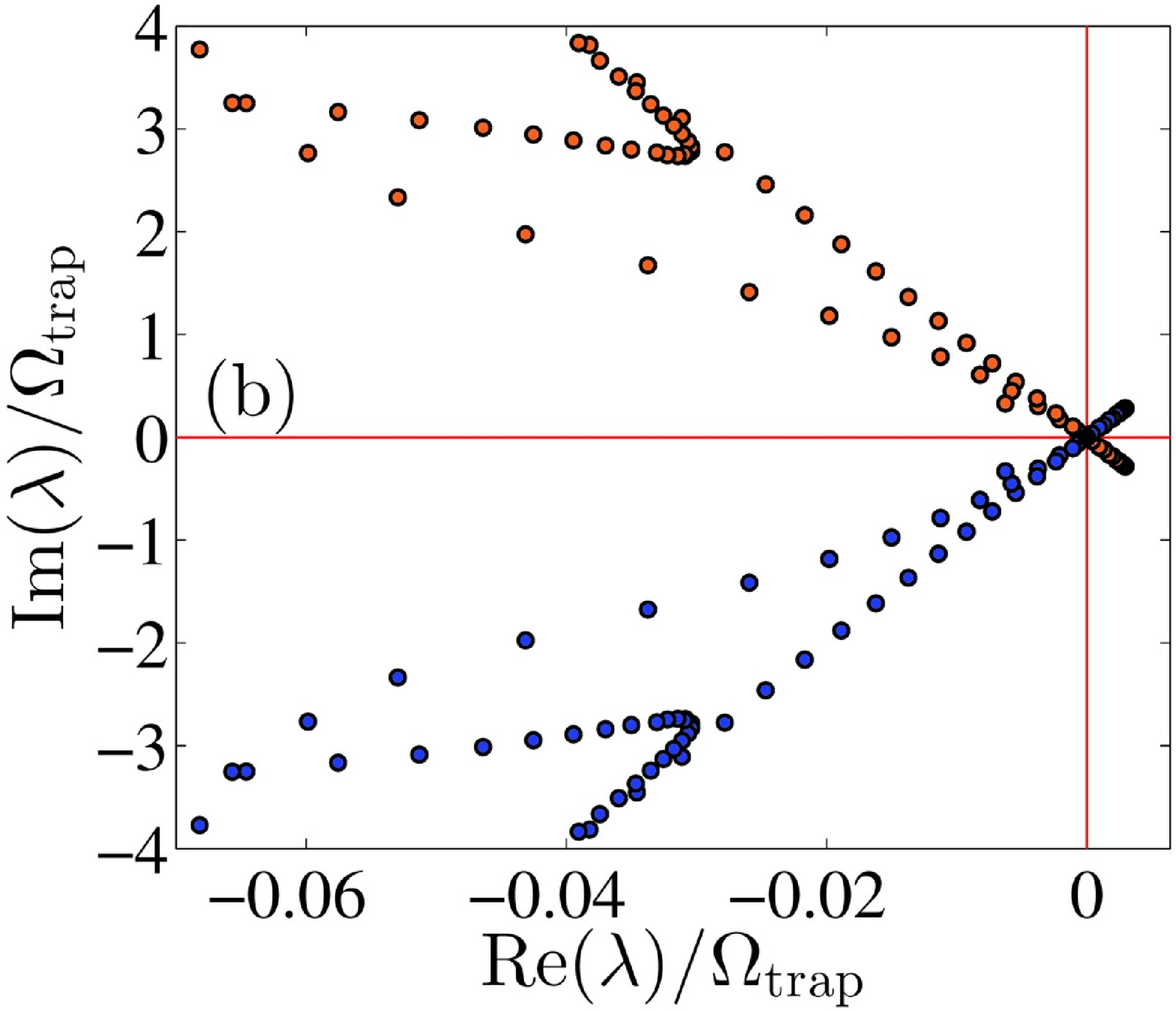}&
\hskip-0.3cm
\includegraphics[height=3.56cm]{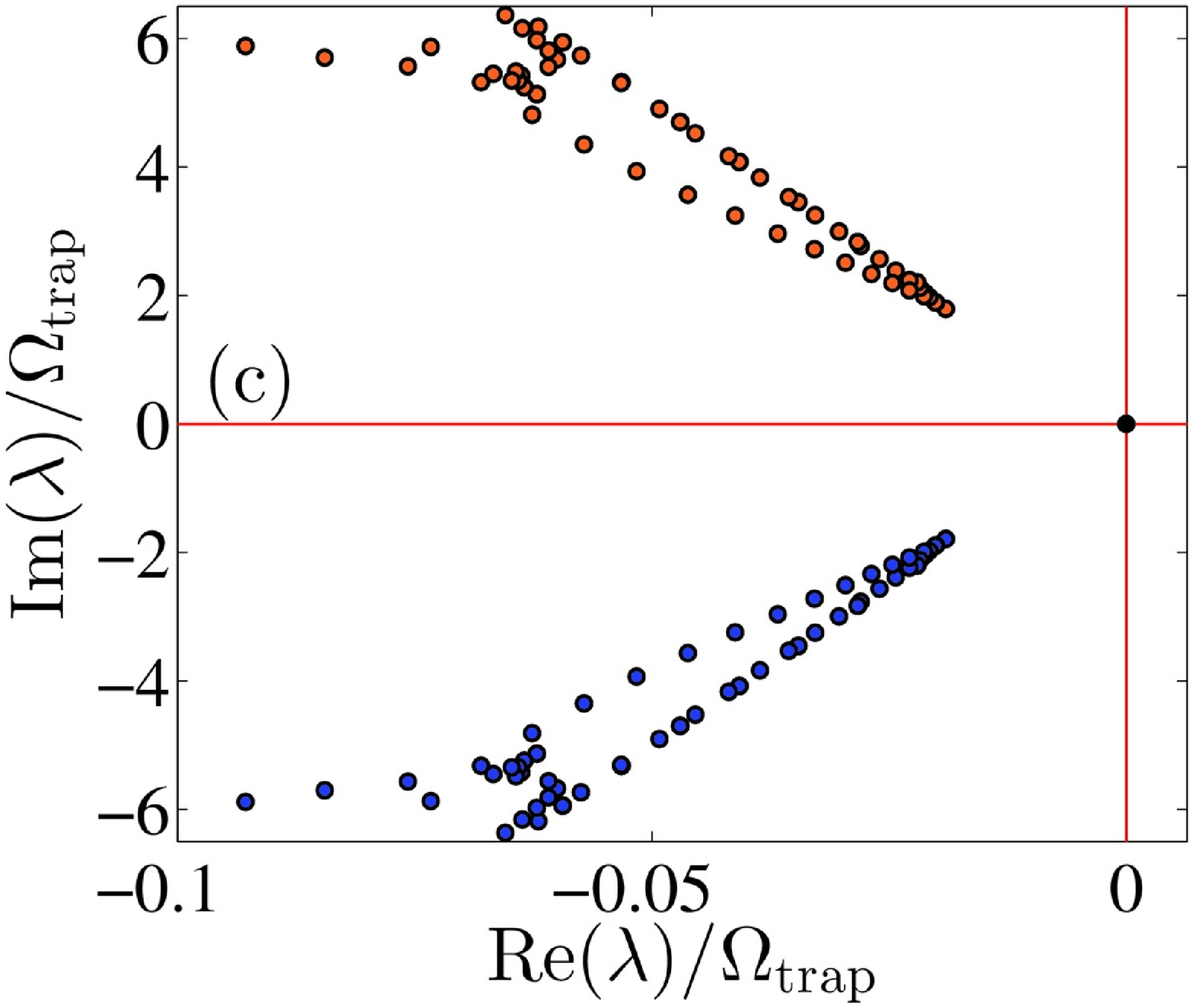}
\\
\hskip-0.1cm
\includegraphics[width=4.10cm]{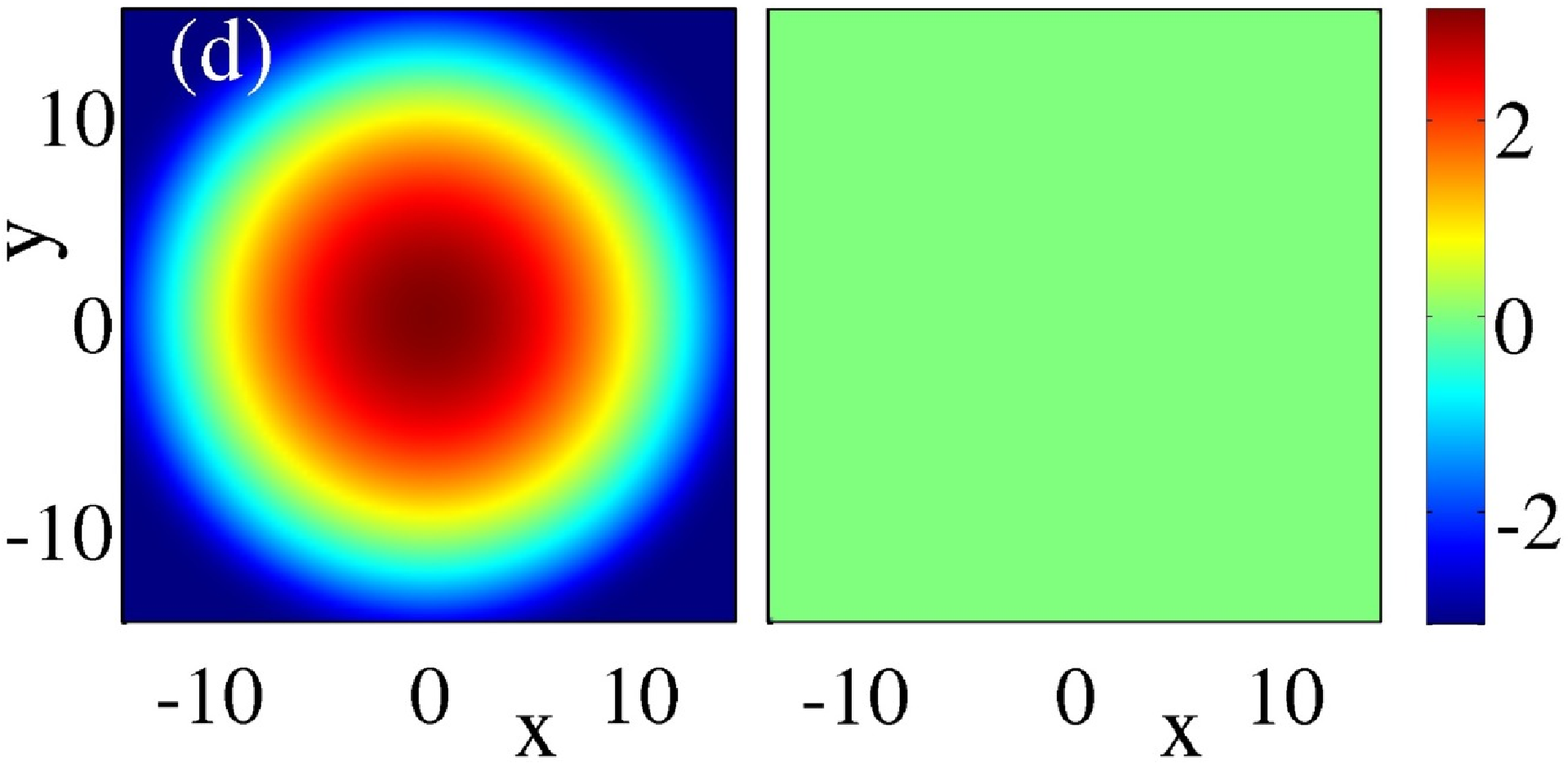}&
\hskip-0.3cm
\includegraphics[width=4.10cm]{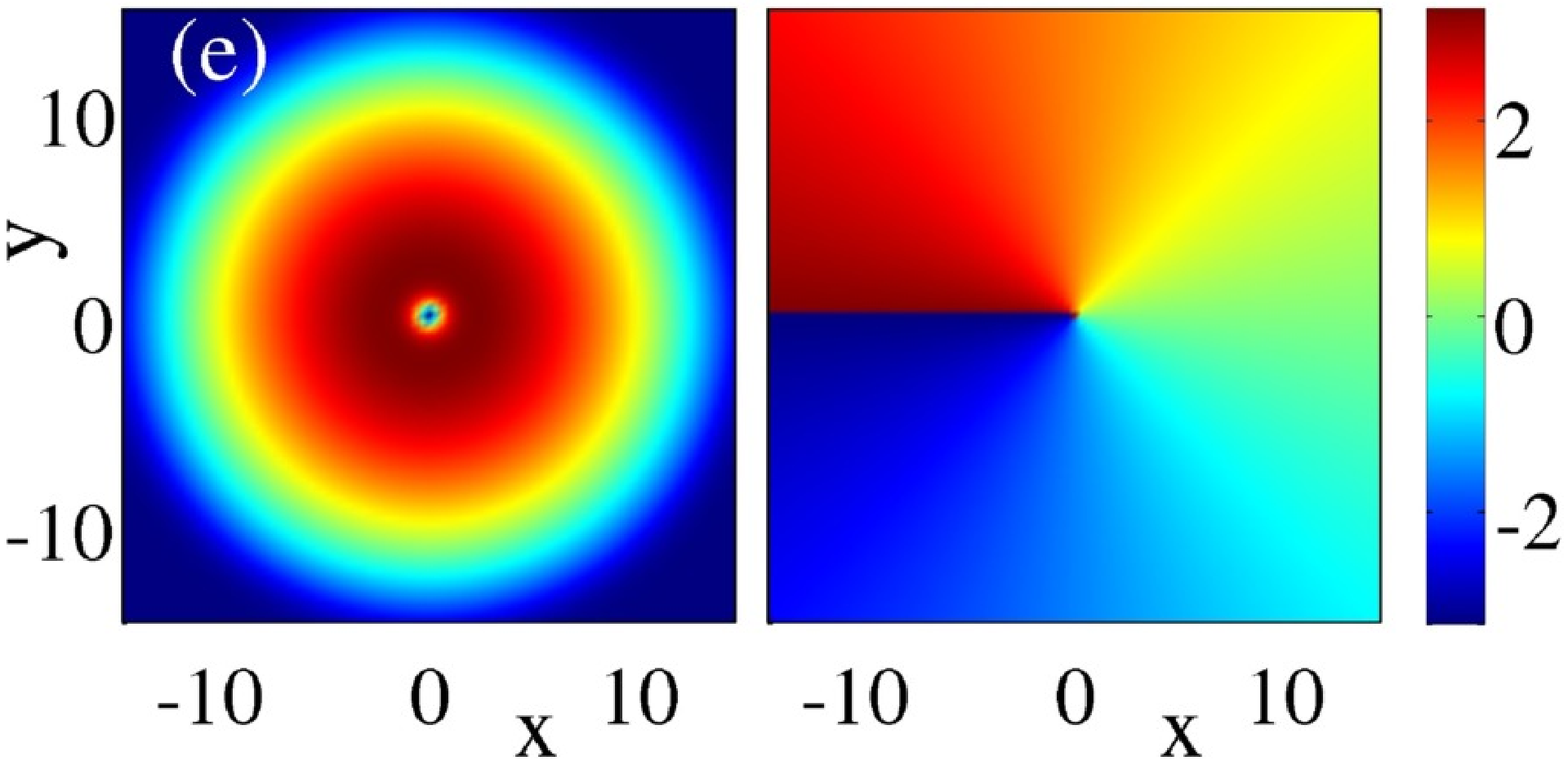}&
\hskip-0.3cm
\includegraphics[width=4.10cm]{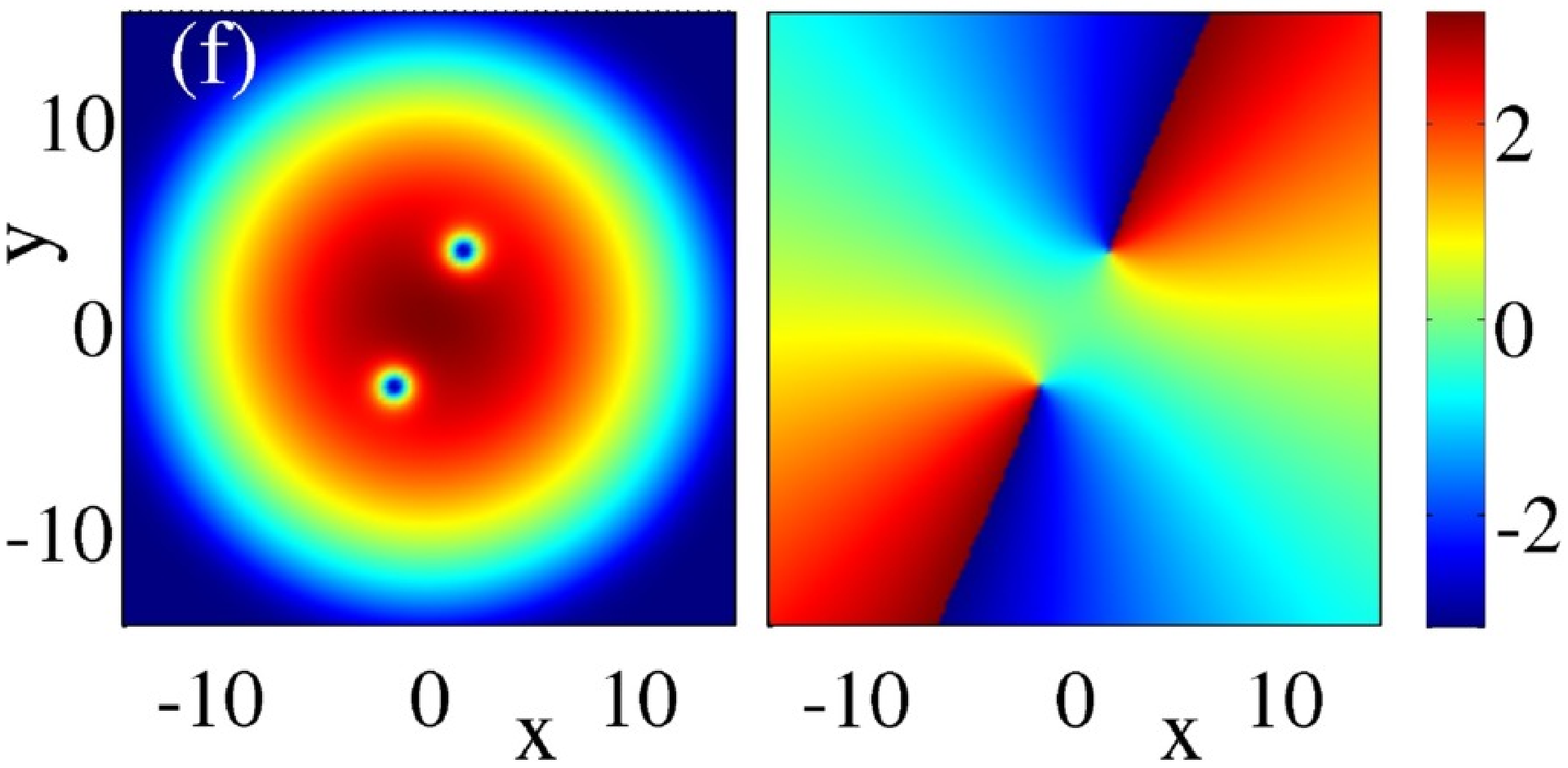}
\end{tabular}
\begin{tabular}{cc}
\hskip-0.1cm
\includegraphics[width=6.25cm]{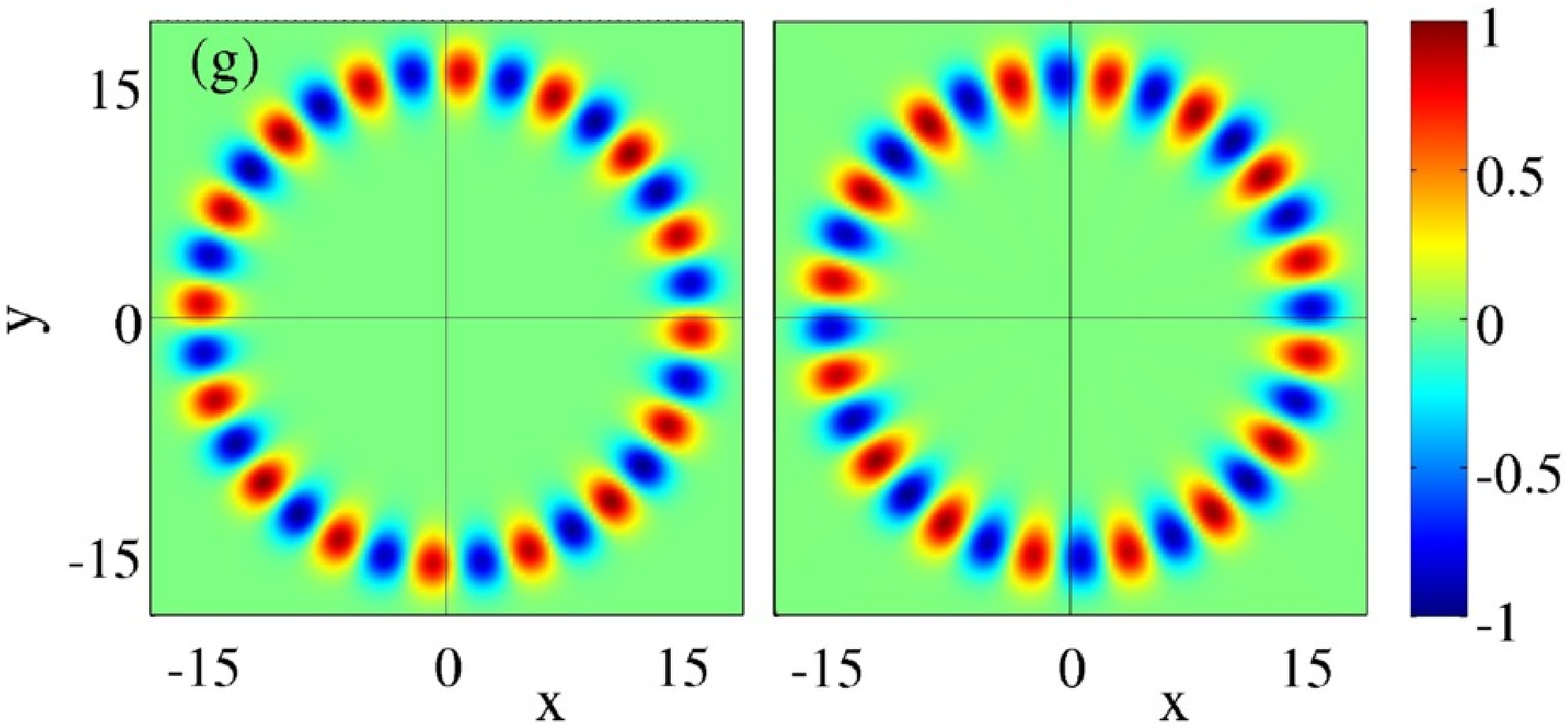}
\hskip-0.3cm
\includegraphics[width=6.25cm]{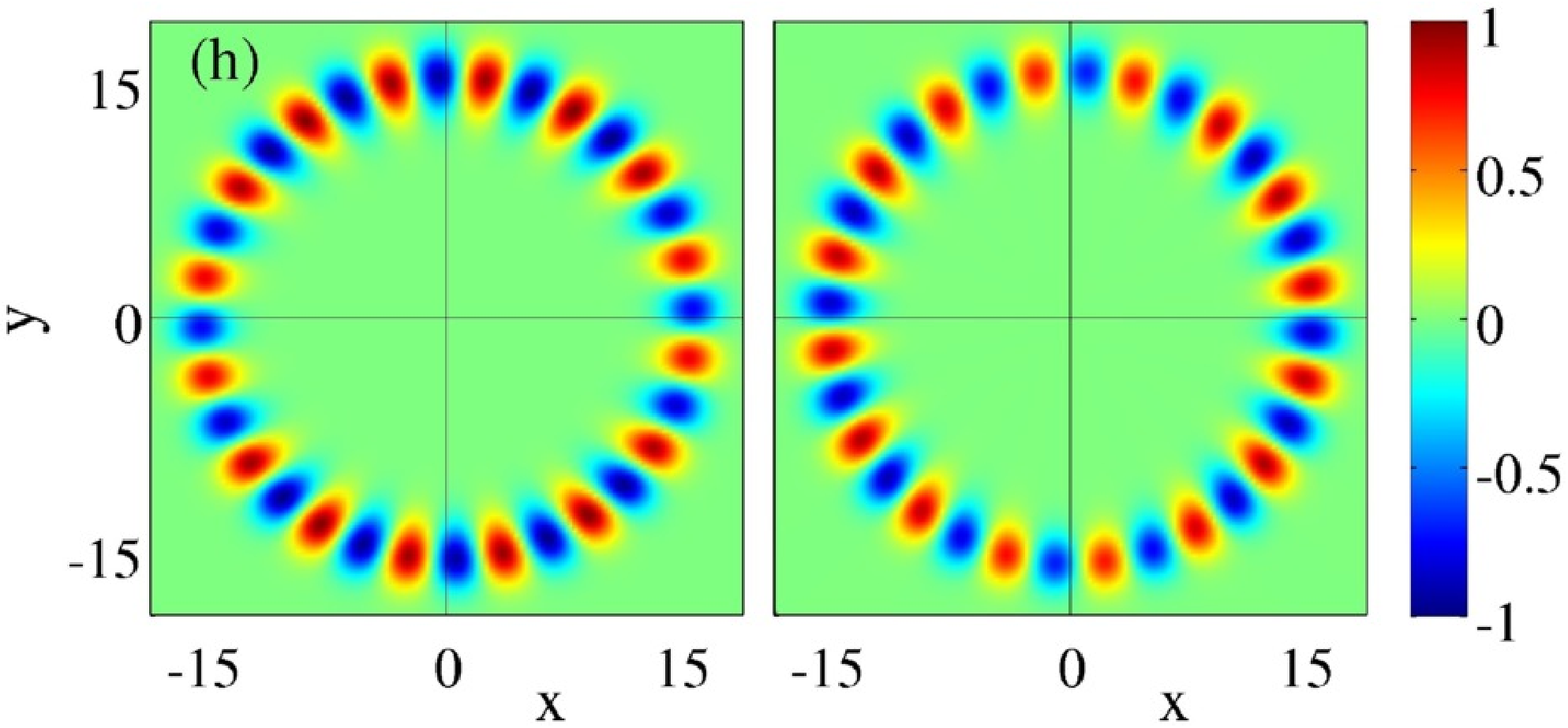}
\end{tabular}
\end{center}
\caption{(color online)
Stability of steady states with different number of vortices for
$\mu=5$, $\gamma=0.01$, $\Omega_{\rm trap}=0.2$, and $\Omega_{\rm rot}/\Omega=0.37$.
(a)--(c) Stability spectra for configurations with 0, 1 and 2
vortices respectively.
(d)--(f) Corresponding density (left subpanels) and phases
(right subpanels) for these configurations.
(g),(h) Most unstable eigenfunctions for configurations with
0 and 1 vortices, respectively.
The left and right subpanels corresponds, respectively, to the
real and imaginary parts of the most unstable eigenfunction.
}
\label{fig:eigenNv}
\end{figure}

\begin{figure}[tb]
\begin{center}
\begin{tabular}{cccc}
\hskip-0.1cm
\includegraphics[height=2.67cm]{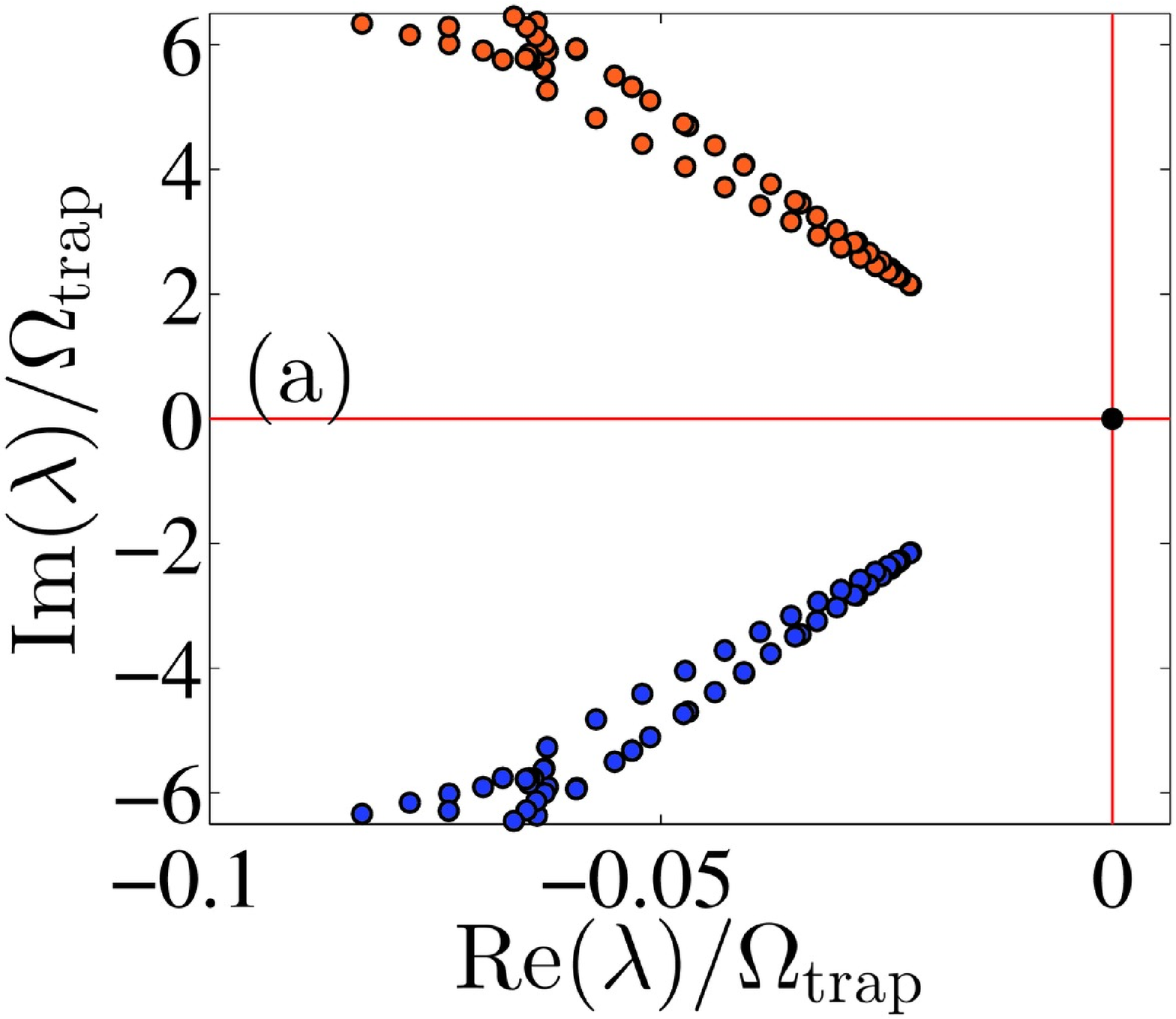}&
\hskip-0.3cm
\includegraphics[height=2.67cm]{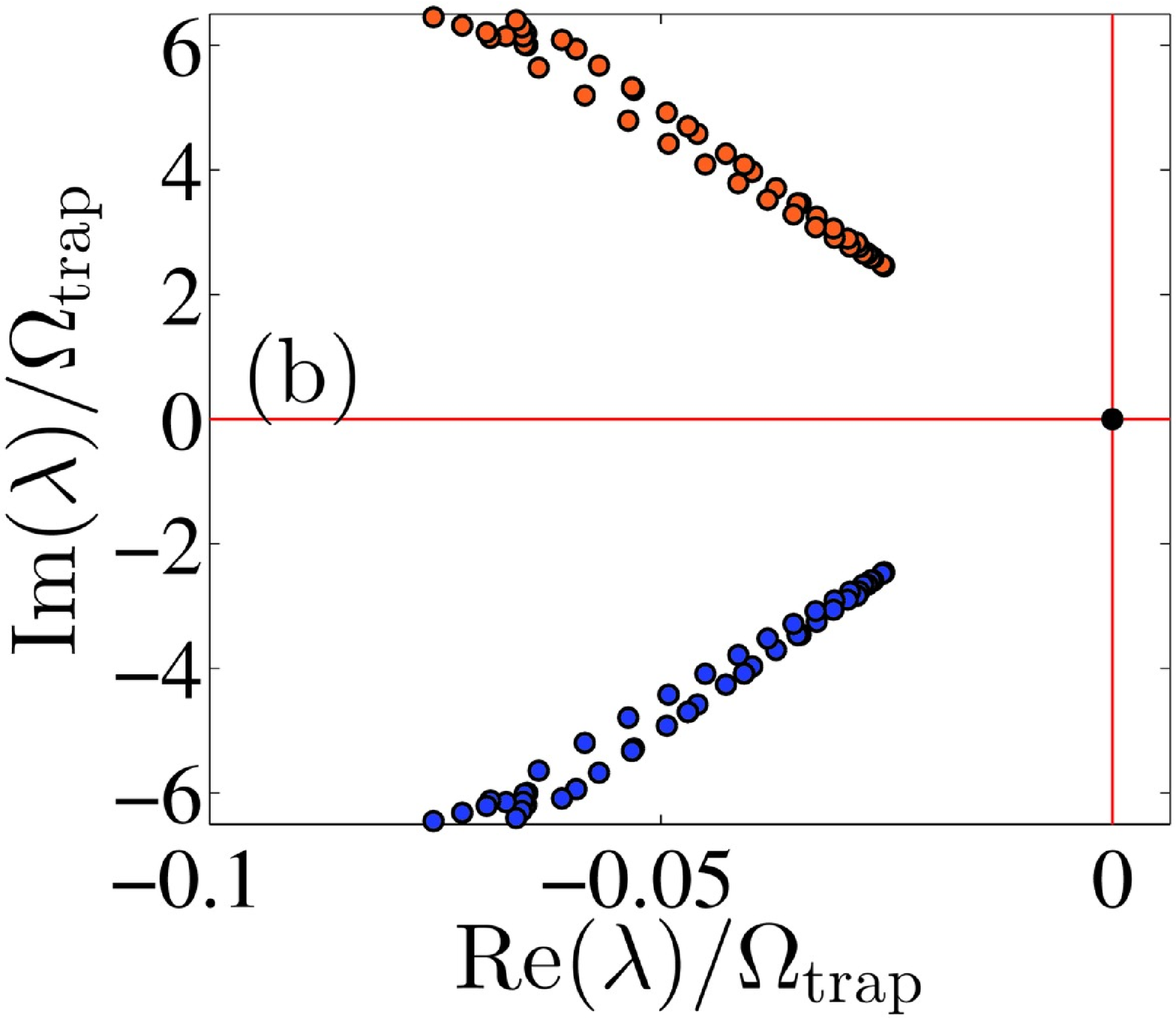}&
\hskip-0.3cm
\includegraphics[height=2.67cm]{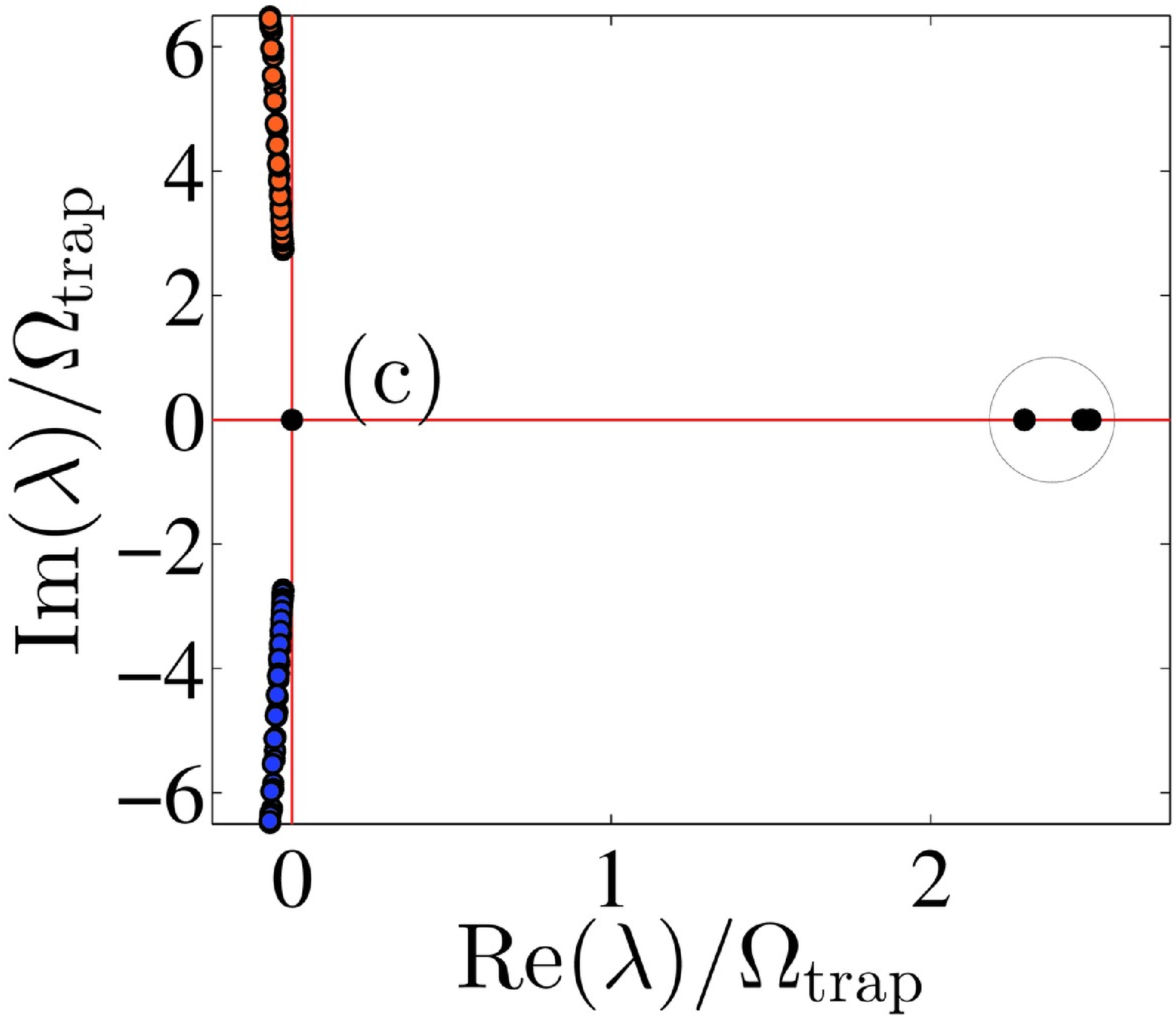}&
\hskip-0.3cm
\includegraphics[height=2.67cm]{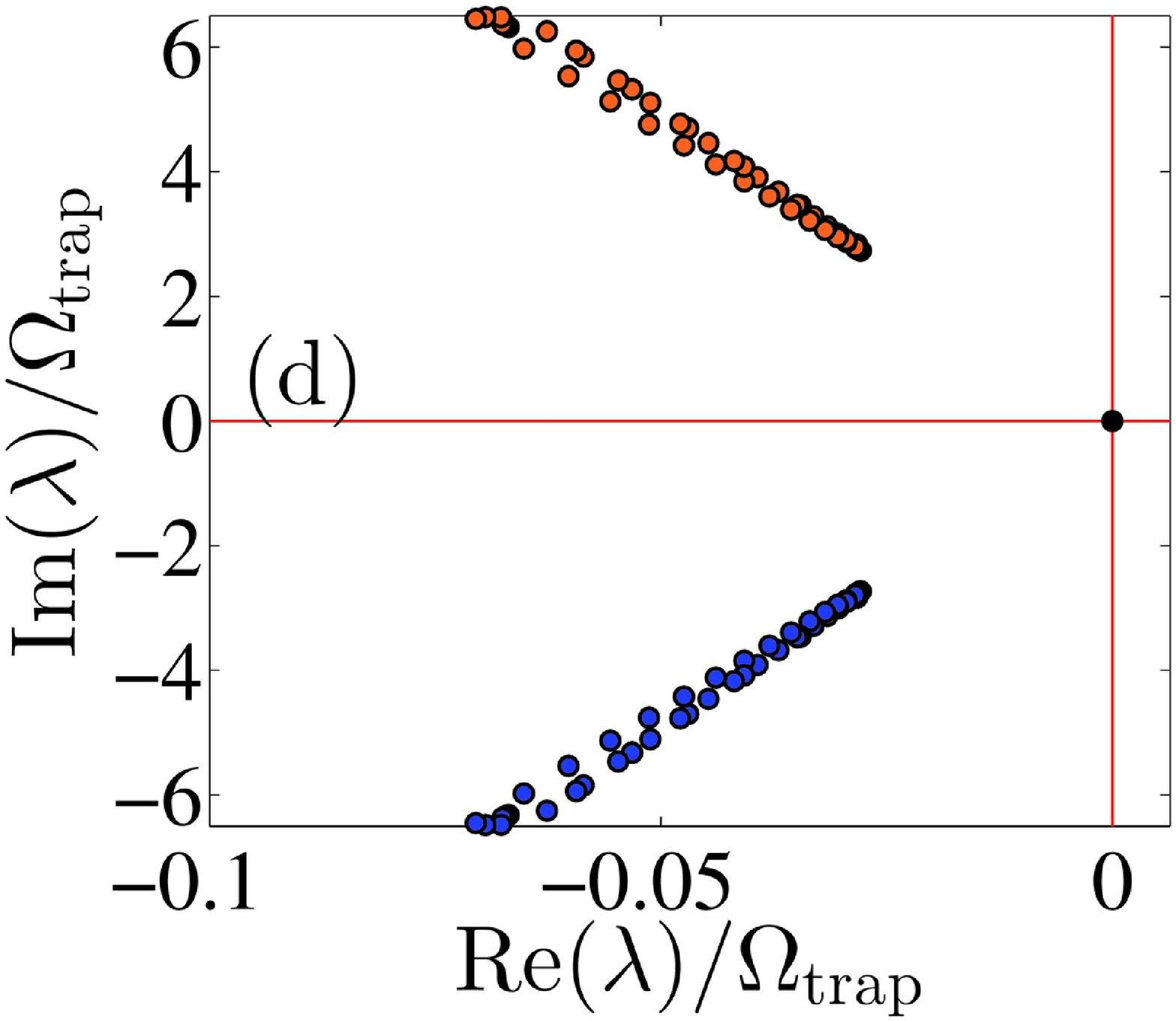}
\end{tabular}
\end{center}
\begin{center}
\begin{tabular}{ccc}
\hskip-0.1cm
\includegraphics[width=4.10cm]{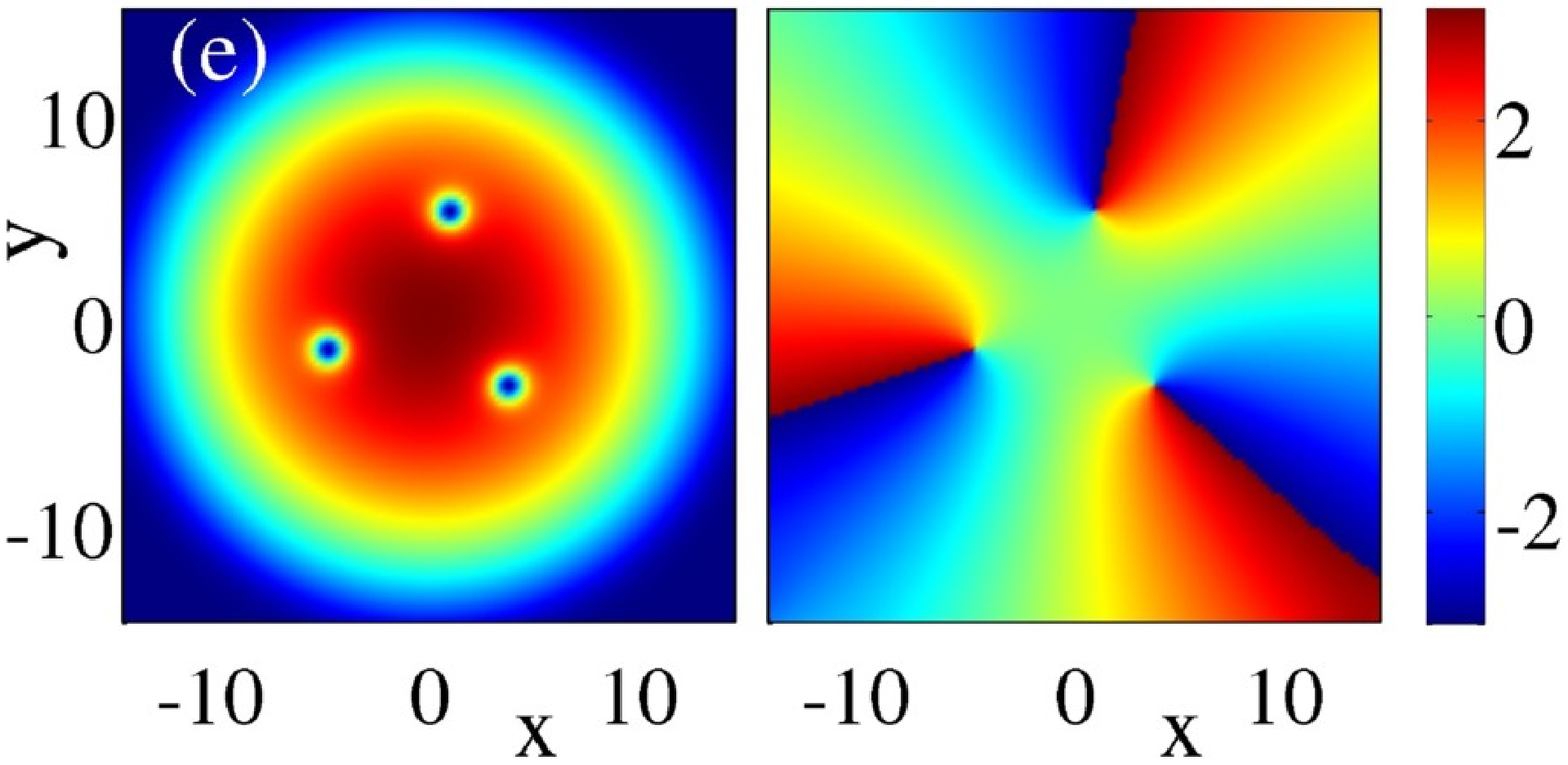}&
\hskip-0.3cm
\includegraphics[width=4.10cm]{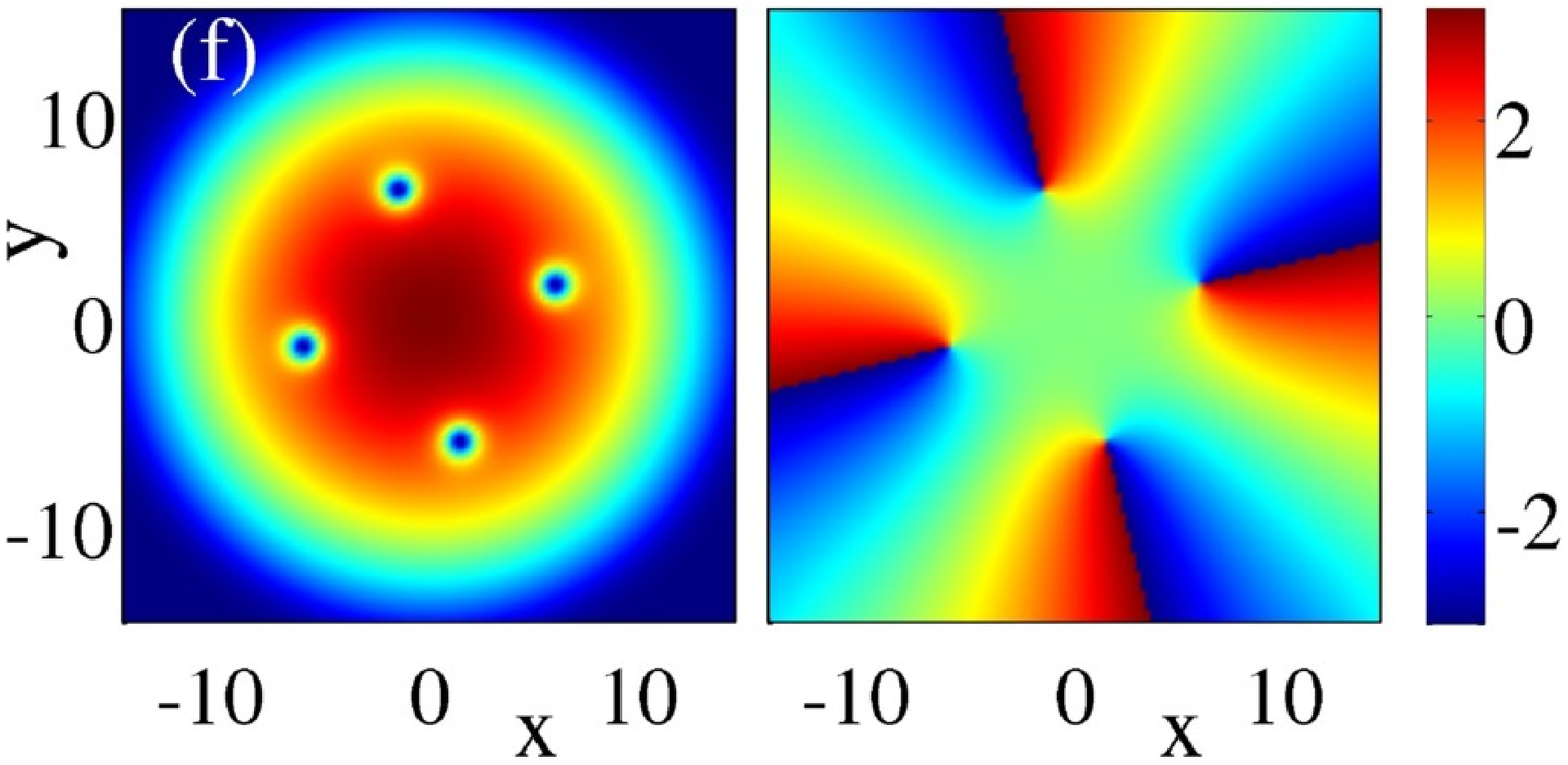}&
\hskip-0.3cm
\includegraphics[width=4.10cm]{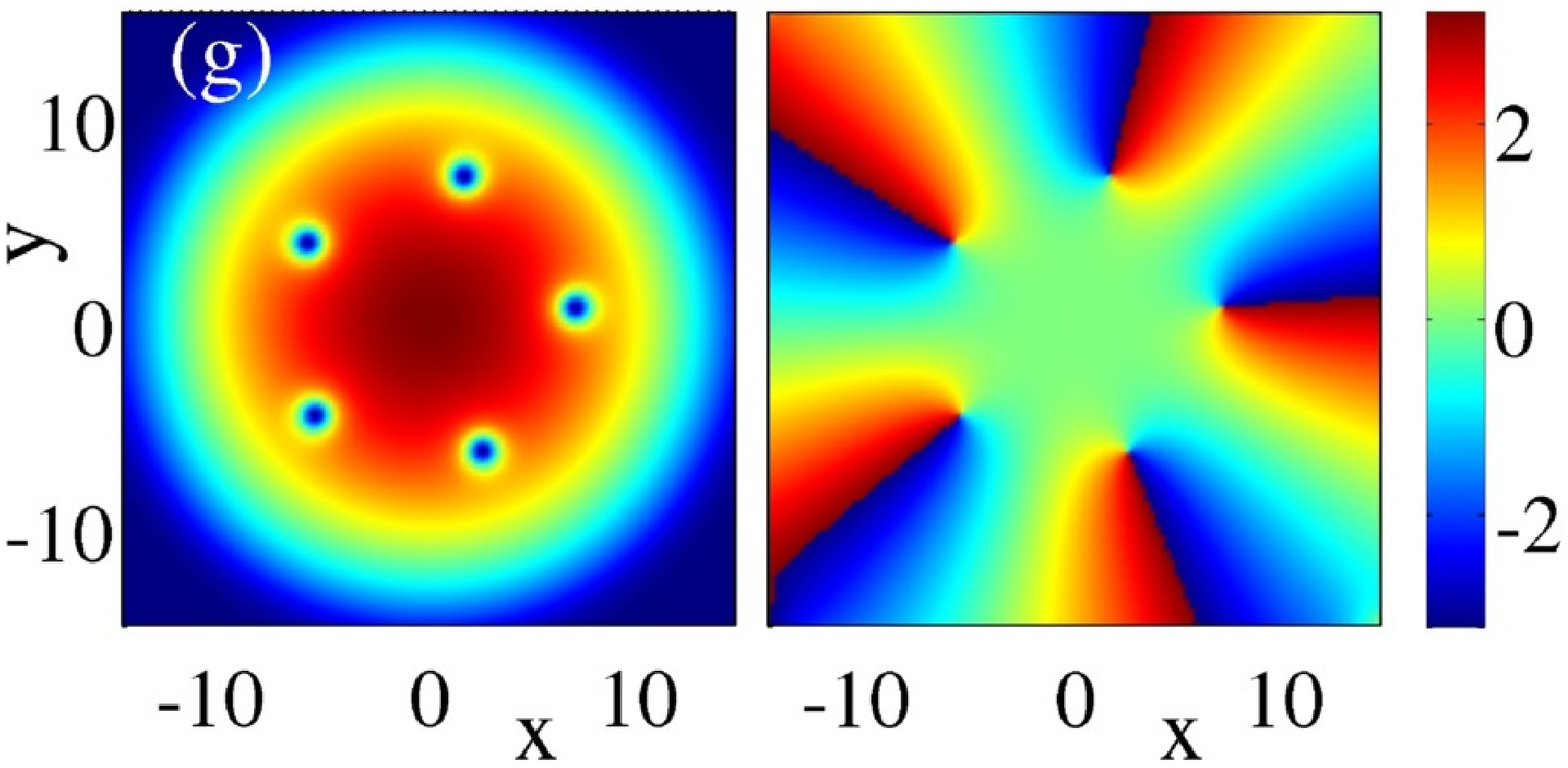}
\\
\hskip-0.1cm
\includegraphics[width=4.10cm]{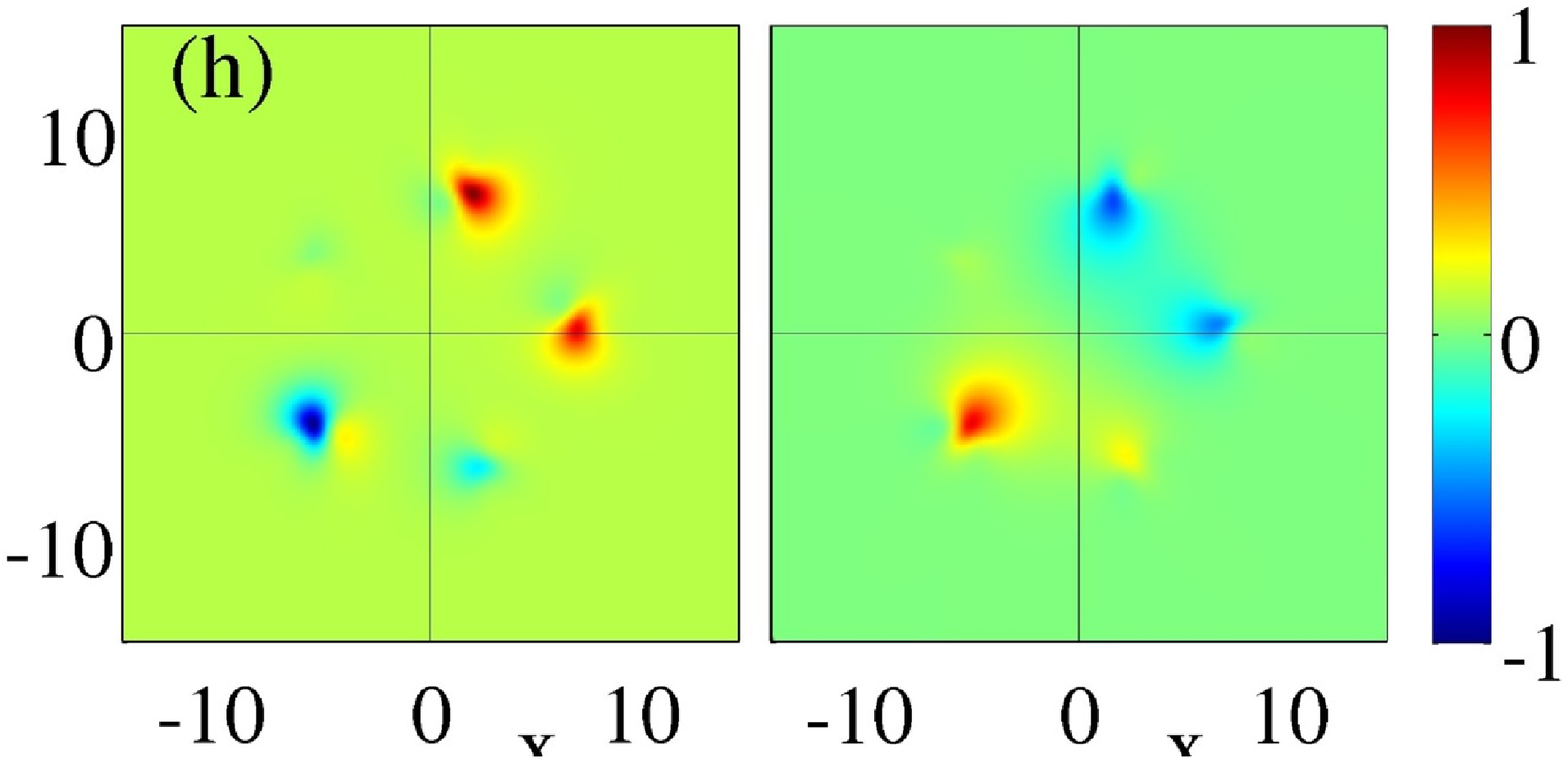}&
\hskip-0.3cm
\includegraphics[width=4.10cm]{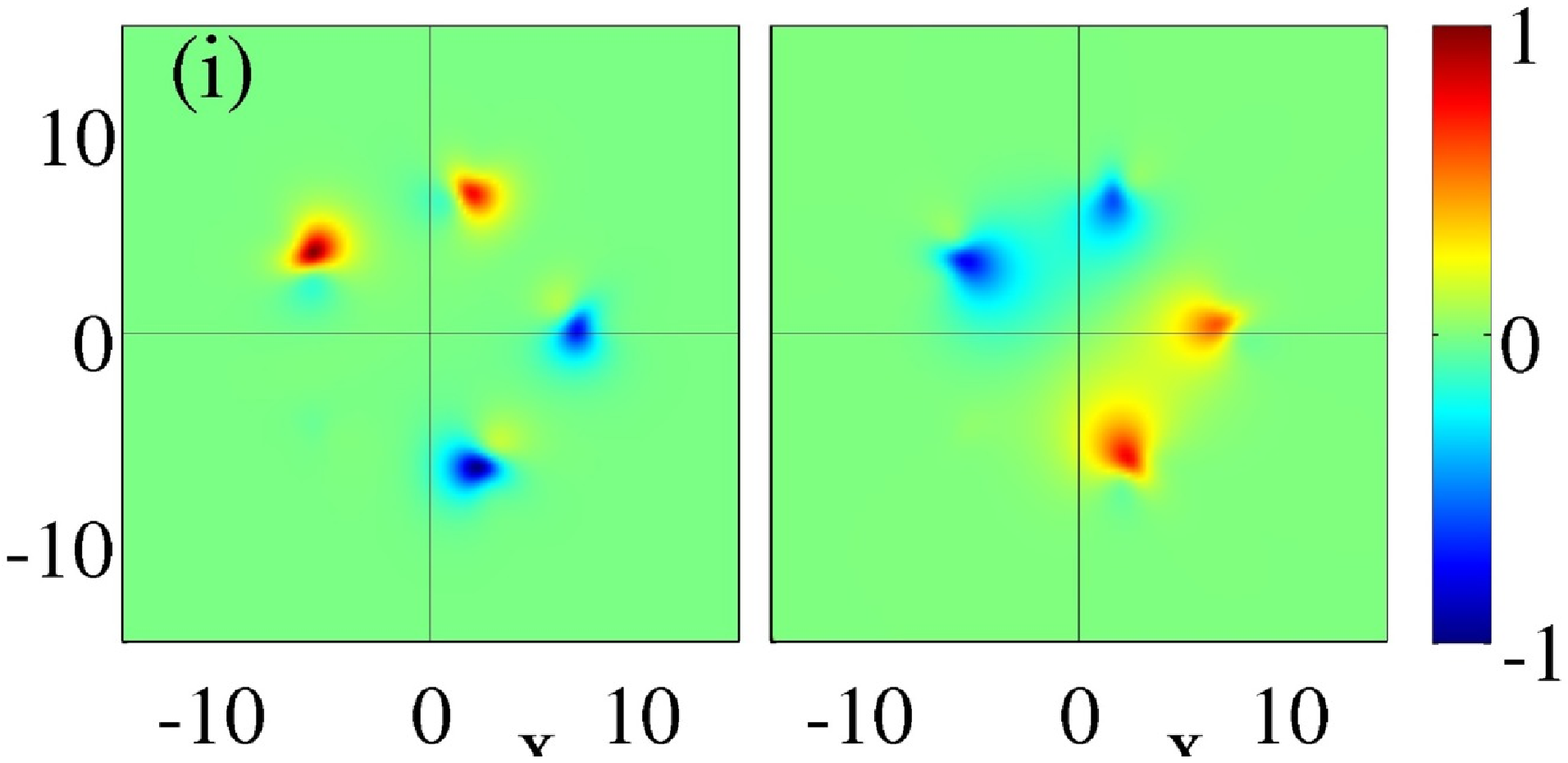}&
\hskip-0.3cm
\includegraphics[width=4.10cm]{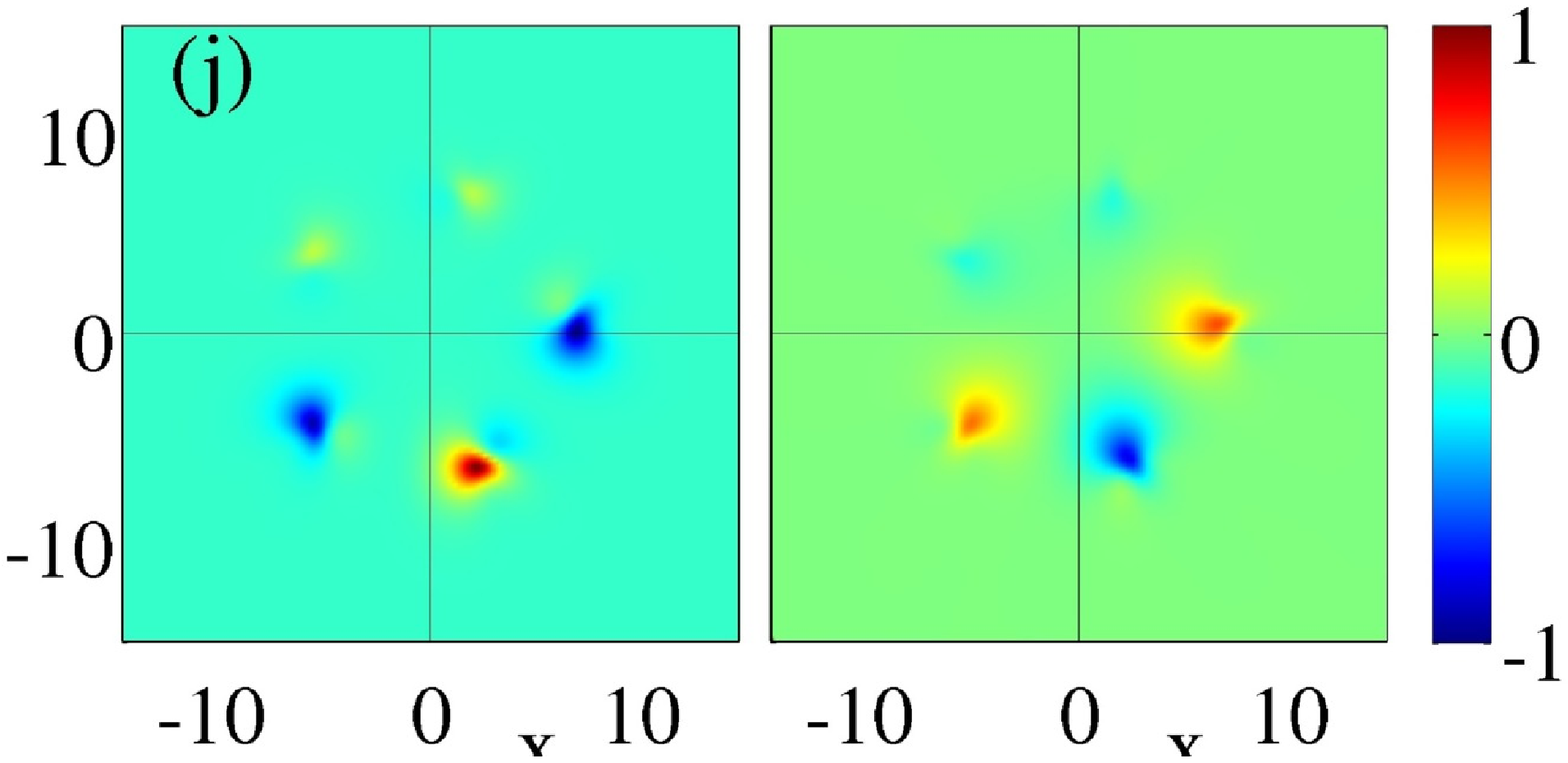}
\end{tabular}
\end{center}
\caption{(color online)
Stability of steady states with different number of vortices for
$\mu=5$, $\gamma=0.01$, $\Omega_{\rm trap}=0.2$, and
$\Omega_{\rm rot}/\Omega_{\rm trap}=0.37$.
(a)--(c) Stability spectra for configurations with 3, 4 and 5
vortices respectively.
Note the unstable eigenvalues inside the circle
for 5 vortices in panel (c).
(d) Zoomed-in version of the stability spectrum for 5 vortices
showing the eigenvalues corresponding to angular modes.
(e)--(g) Density (left subpanels) and phases
(right subpanels) for the configurations with 3, 4 and 5
vortices respectively.
(h)--(j) Three most unstable eigenfunctions for the configuration with
5 vortices.
The left and right subpanels corresponds, respectively, to the
real and imaginary parts of the most unstable eigenfunction.
}
\label{fig:eigenNv345}
\end{figure}

The final fate of the above selection mechanism can
be, at least in part, be attributed to the fact that stationary
corotating vortex polygons with different numbers of vortices
have different stability properties
for fixed parameter values.
For instance, in Fig.~\ref{fig:eigenNv} we depict the steady
states, their stability spectra and their most unstable eigenmode
for the case of $\mu=5$ for zero, one and two central vortices.
As it is evident from the figure, the configurations bearing zero
vortices and one vortex are unstable while the configuration with
two vortices is stable.
This corroborates the dynamical evolution depicted in
Fig.~\ref{fig:snap_film_mu05} where the initial state with
zero vortices destabilizes towards a transient state
with one vortex that, in turn, destabilizes towards the
final, stable steady state with two vortices.
A similar stability analysis for $\mu=10$ (results not shown here)
indicates that indeed the steady states with zero, one and two vortices
are all unstable, while the state with three is perfectly stable;
corroborating what is observed in the dynamical evolution
depicted in Fig.~\ref{fig:snap_film_mu10}.

\begin{figure}[ht]
\begin{center}
\includegraphics[width=12cm]{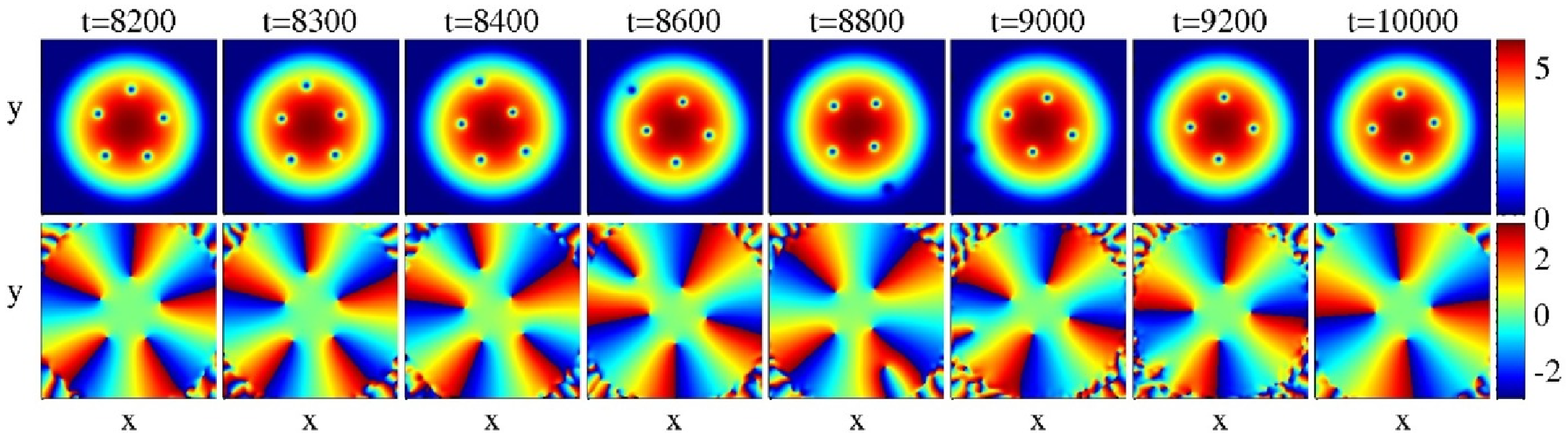}
\end{center}
\caption{(color online)
Evolution of an unstable 5-vortex configuration.
The initial state corresponds to the one in Fig.~\ref{fig:eigenNv345}(g)
(namely $\Omega_{\rm rot}/\Omega_{\rm trap}=0.37$, $\mu=5$, $\gamma=0.01$
and $\Omega_{\rm trap}=0.2$).
The windows for density and phase are,
respectively,  $(x,y)\in[-17.5,17.5]\times[-17.5,17.5]$ and
$(x,y)\in[-19,19]\times[-19,19]$.
We invite the interested reader to see the full movie at this address:
\href{http://nonlinear.sdsu.edu/~carreter/RotatingBEC.html}{http:$\sslash$nonlinear.sdsu.edu/~carreter/RotatingBEC.html} [Movie\#3].
}
\label{fig:snap_film_mu05_Nv5}
\end{figure}

The above results prompt the question of stability for
configurations bearing an increasing number of vortices.
For instance, as previously described, the configuration without vortices
destabilizes when the rotation is increased, while configurations
with one or more vortices become stable.
However, polygonal configurations have a limit to the
number of vortices that they can hold before becoming unstable
(see Ref.~\cite{theo14} and references therein).
This is depicted in Fig.~\ref{fig:eigenNv345} where the stability
for polygonal vortex states with 3, 4 and 5 vortices
for the case $\mu=5$ is examined.
As it is clear from these results, the polygonal configurations with
3 and 4 vortices are indeed stable while the one with 5 vortices
is unstable.
Interestingly, the instability responsible for the breakup of
the 5-vortex configuration is {\em not} an angular mode as in
all the cases presented previously. In fact, it is clear that
all the angular modes are stable as it can be seen from the
zoomed-in version of the spectrum depicted in panel (d) of the figure
where only the eigenvalues associated with angular modes are
depicted.
Nonetheless, the 5-vortex state is indeed unstable as is
evident by the small cluster of unstable eigenvalues enclosed
in the small circle in panel (c) of the figure. The eigenmodes
associated with these unstable eigenvalues are depicted in
panels (h)--(j). These symmetry breaking
modes bring some of the vortices closer and others further apart
from each other.
It is precisely this mechanism that is responsible for the
destabilization of the 5-vortex polygonal state as it is depicted
in the snapshots of its evolution in Fig.~\ref{fig:snap_film_mu05_Nv5}.
We invite the interested reader to see the full movie at this address:
\href{http://nonlinear.sdsu.edu/~carreter/RotatingBEC.html}{http:$\sslash$nonlinear.sdsu.edu/~carreter/RotatingBEC.html} [Movie\#3].
Here, the initial steady state bearing a
5-vortex polygonal state destabilizes around $t=8200$,
via a mode that pushes some vortices inward and other outward,
eventually resulting in a {\em stable} 4-vortex polygonal state after one vortex
is ejected towards the periphery of the cloud.

\begin{figure}[ht]
\begin{center}
\includegraphics[width=4.1cm]{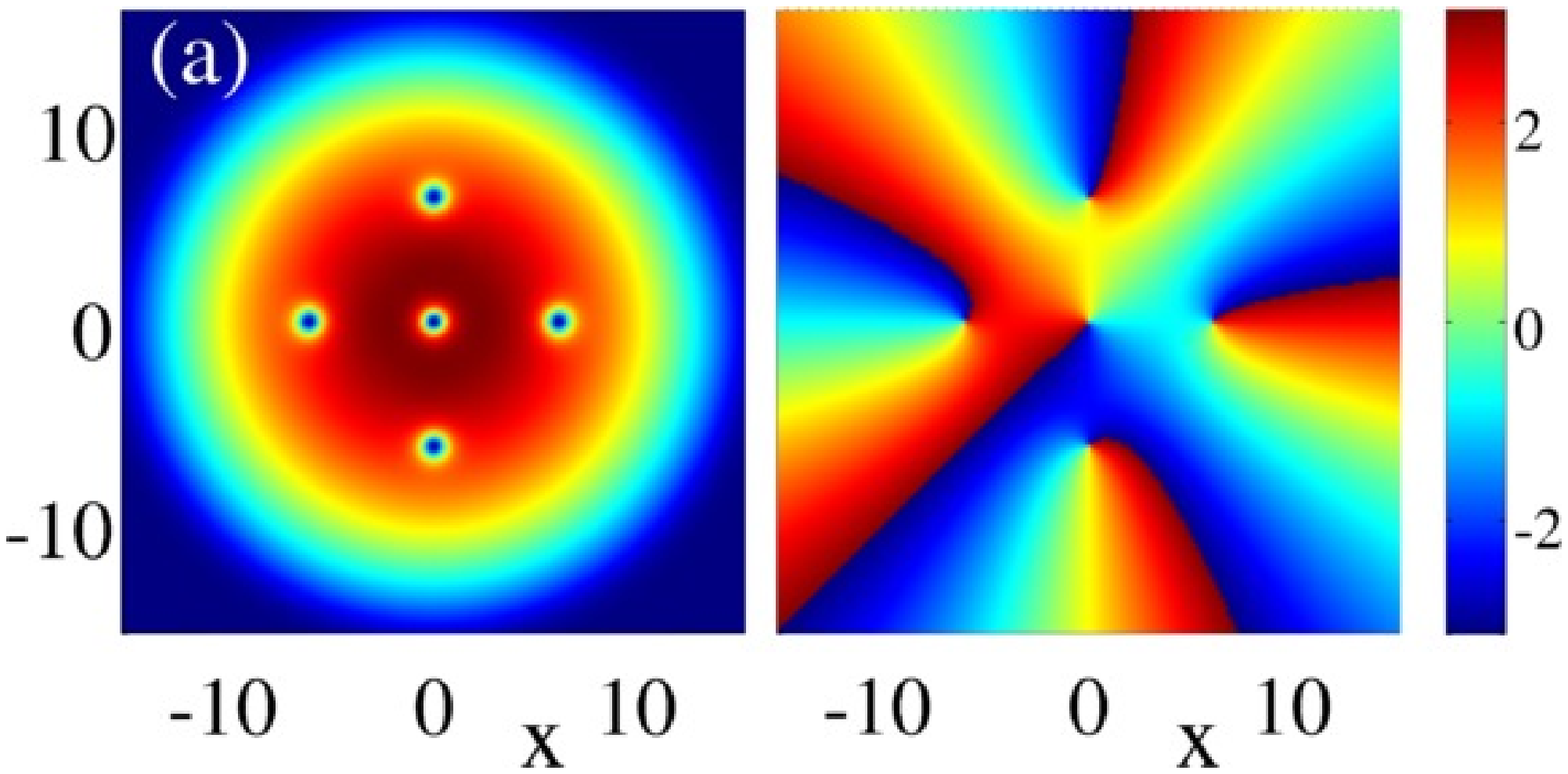}
\includegraphics[width=4.1cm]{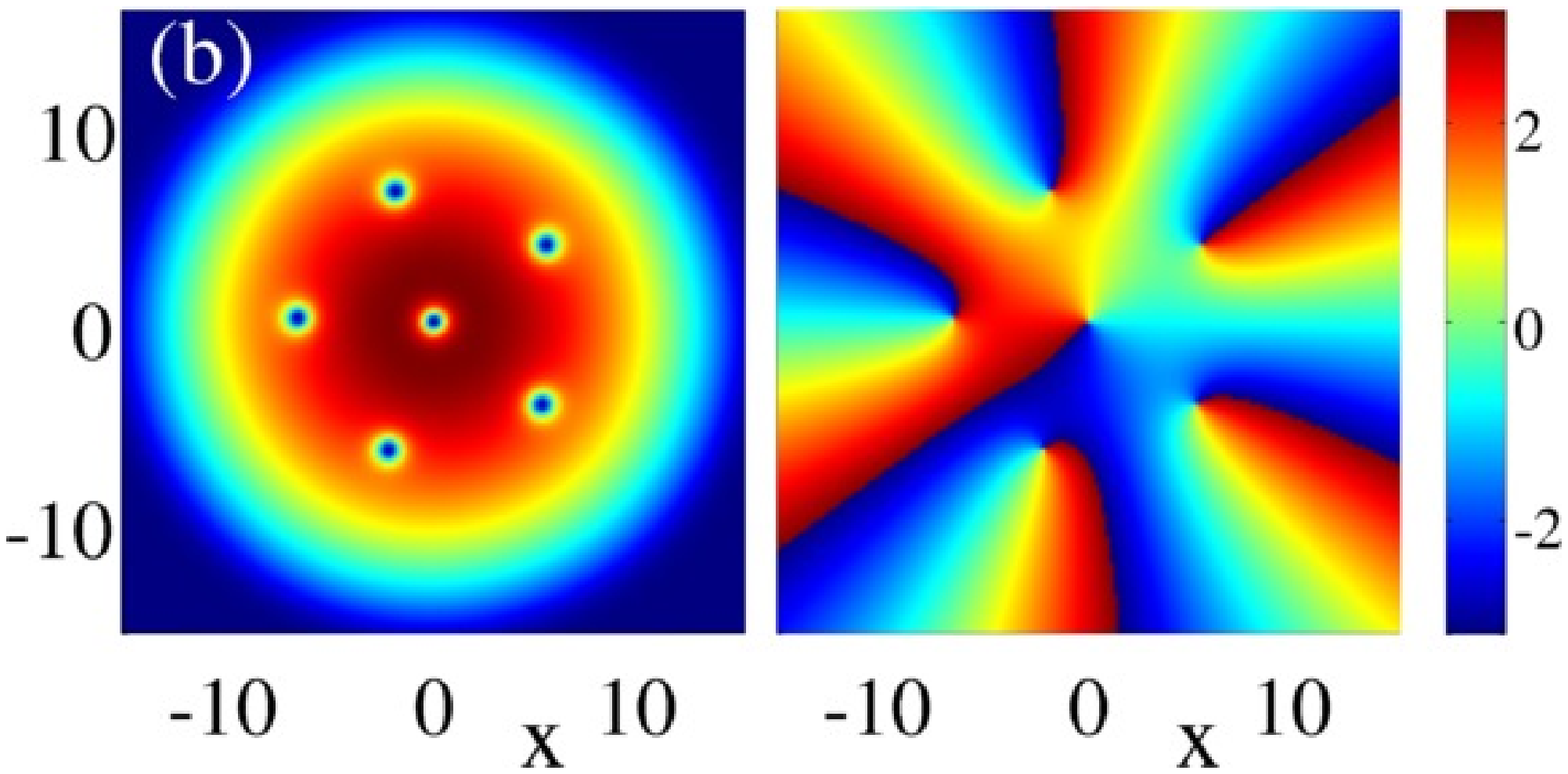}
\includegraphics[width=4.1cm]{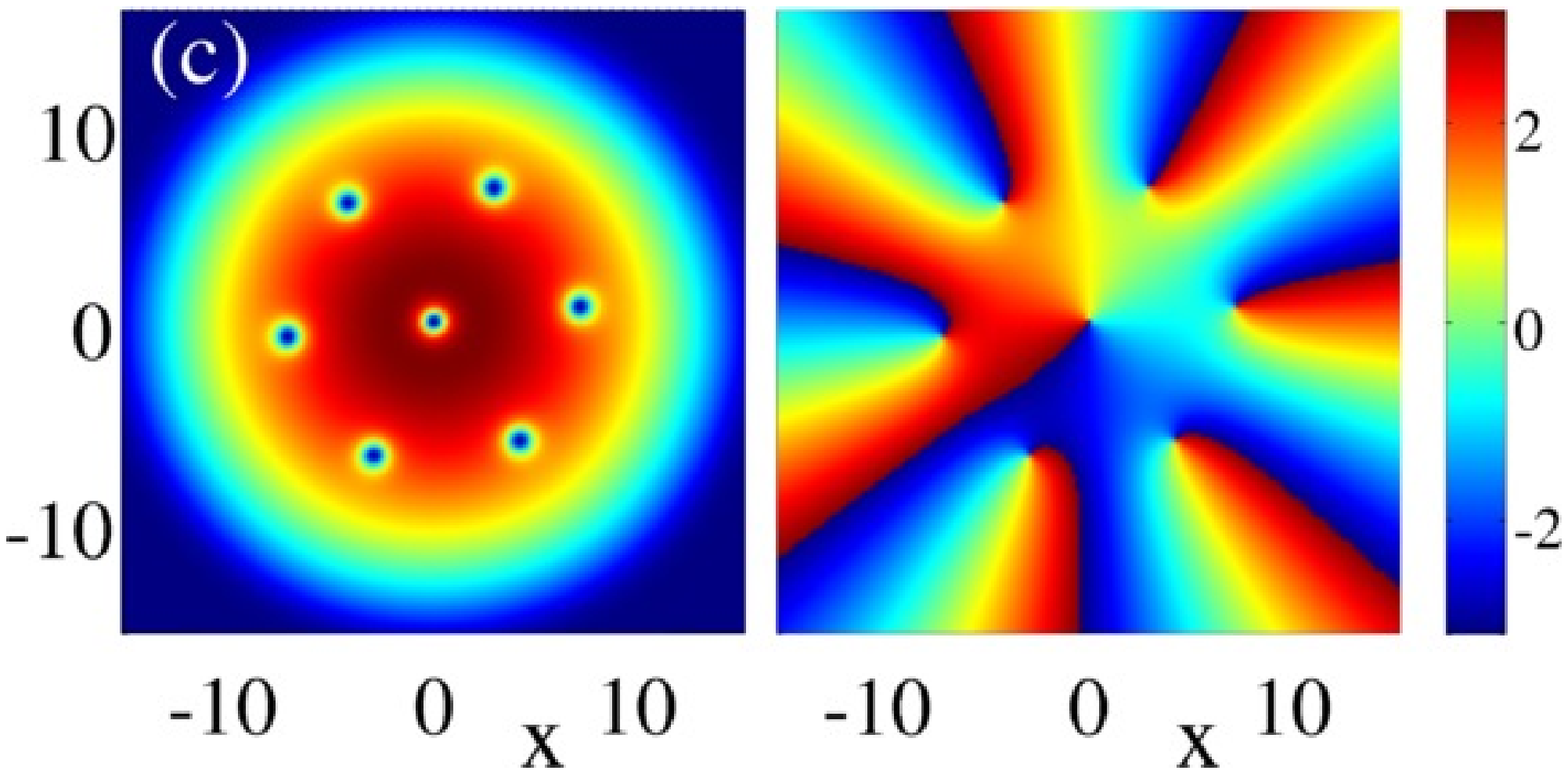}
\end{center}
\caption{(color online)
Steady state configurations with $N$+1 vortices for
$\Omega_{\rm rot}/\Omega_{\rm trap}=0.37$, $\mu=5$, $\gamma=0.01$,
and $\Omega_{\rm trap}=0.2$.
(a) 4+1 configuration,
(b) 5+1 configuration, and
(c) 6+1 configuration.
All these configurations are stable for the chosen parameter values.
}
\label{fig:Nv+1}
\end{figure}

We have checked that the above phenomenology persists for other values
of the dissipation coefficient $\gamma$. For instance, although the stability 
thresholds and the order ($m$) of the unstable modes for the vortex-less 
configuration are {\em independent} of $\gamma$, the growth rates for these 
instabilities are indeed dependent on dissipation (see also discussion below 
in Sec.~\ref{sec:Model}). In particular, the larger $\gamma$ is, 
the larger
the instability growth rates will be. Therefore, for larger values of $\gamma$ the
instability will set in earlier and, more importantly, the pattern selection
mechanism for the setlling of a cluster of vortices at the center of the
cloud will be different. This is a direct consequence from the fact that
the spiraling experienced by a vortex due to dissipation has a faster
radial rate as $\gamma$ is increased \cite{dongyan}.
A key effect of the slow down in the radial spiraling rate
as $\gamma$ is decreased is that the pattern selection mechanism
has ``more time'' to select votices and thus a smaller number
of vortices is eventually pulled in from the periphery.
This is precisely what we observe in numerical simulations where smaller
values of $\gamma$ give rise to final configurations with a smaller
number of vortices (results not shown here).
For instance, for $\mu=5$ we find that for values of $\gamma$ of
$0.5$, $0.2$, $0.1$, and $0.01,$
a vortex-less configuration evolves
towards a stable steady state configuration with, respectively,
9, 7, 3, and 2
vortices.
The same setup but for $\mu=10$ yields, respectively,
15, 11, 5, and 3
vortices.
We invite the interested reader to see the full movies
for all of these cases at this address:
\href{http://nonlinear.sdsu.edu/~carreter/RotatingBEC.html}{http:$\sslash$nonlinear.sdsu.edu/~carreter/RotatingBEC.html} 
[Movies\#4--11].

Finally, let us briefly touch upon the existence of other
relevant vortex configurations. The fact that polygonal
configurations with large number of vortices lose stability
prompts the important question: what are the remaining stable configurations
of the system? For instance, in the absence of rotation,
it has been shown that polygonal configurations 
become destabilized towards asymmetric configurations in a symmetry-breaking
pitchfork bifurcation~\cite{zampetaki,theo14} that has been
observed in actual BEC experiments~\cite{dshall3}.
This instability occurs when the polygonal configuration increases
its radius, namely, when the angular momentum of the vortex
cluster is increased.
In a similar manner, as we increase the rotation in the
DGPE model~(\ref{eq:DGPE}) or as we increase the number of
vortices, polygonal configurations lose stability.
These instabilities can be manifested through angular modes
(cf.~the case of one vortex in Fig.~\ref{fig:eigenNv})
or through symmetry breaking modes (cf.~the case for 5 vortices
in Figs.~\ref{fig:eigenNv345} and \ref{fig:snap_film_mu05_Nv5}).
However, on the other hand, if ones starts with a polygonal
state with an extra vortex at the center, the so-called
$N$+1 vortex configurations, new stable states are
produced~\cite{theo14} that might even be the ground states
(i.e., the minimal energy states)
of the system~\cite{zampetaki}.
In Fig.~\ref{fig:Nv+1} we depict three examples of these $N$+1
configurations for 4+1, 5+1, and 6+1 vortices.
It is important to mention that these three configuration
are indeed {\em stable} for the chosen parameter values.
In fact, as the rotation increases and the number of vortices
increases as well, more complex configurations relating to
triangular (Abrikosov) vortex lattices arise.

Although our principal emphasis in this work lies in the identification
of the most unstable mode that results in the instability of the 
state bearing no vortices in the context of Eq.~(\ref{eq:DGPE}), 
clearly numerous additional issues emerge from the above simulations.
These concern the identification of the minimal energy state and
the transition pathways resulting in different types of local or global
attractors. We will briefly return to these questions in the context
of future challenges in Section~\ref{sec:conclu}.

\section{Overdamped NLS Model}
\label{sec:overdampedNLS}

\subsection{Model and Numerical Results}
\label{sec:Model}

Now a crucial observation allows us to explore the principal question
raised above (about the dominant instability mode) in the context of
slightly different and somewhat simpler 
model. The topological nature of the observed
stability characteristics (i.e., of positive and negative energy modes)
of the stationary state without vortices
renders them robust and {\em independent} of the precise value of $\gamma$.
In particular, for different values of $\gamma$,
the actual size of the growth rates
will change (in fact, the higher $\gamma$ is, the more unstable the relevant
modes are). However, as can be also numerically checked, the ordering of the
relevant eigenmodes will not be modified by the precise value of $\gamma$.
That is to say, the same mode will become unstable at the same critical
point (but with a different slope of Re$(\lambda)$
vs.~$\Omega_{\rm rot}/\Omega_{\rm trap}$
in Fig.~\ref{tkfig4}) for a different value of $\gamma$.
Given this feature, we will hereafter choose to explore the
``overdamped'' limit of large $\gamma$, where in fact the Hamiltonian term
in the left hand side of Eq.~(\ref{eq:DGPE}) is neglected in comparison to
the $\gamma$-dependent dissipative one (and subsequently a time rescaling to
absorb $\gamma$ is performed). Hence, we will work below with the
``imaginary time'' variant of the equation
\begin{eqnarray}
u_t=\frac{1}{2} \Delta u - \frac{1}{2} \Omega_{\rm trap}^2 x^2 u + \mu
u - |u|^2 u - i \Omega_{\rm rot} u_{\theta},
\label{tk_eqn4}
\end{eqnarray}
which, based on the above arguments, should be sufficient to provide us with
a prediction for the above instability when focusing
on the vortex-less stationary state. In fact, this very statement will be
double checked a posteriori in the next section.

\begin{figure}[tb]
\begin{center}
\includegraphics[width=0.97\textwidth]{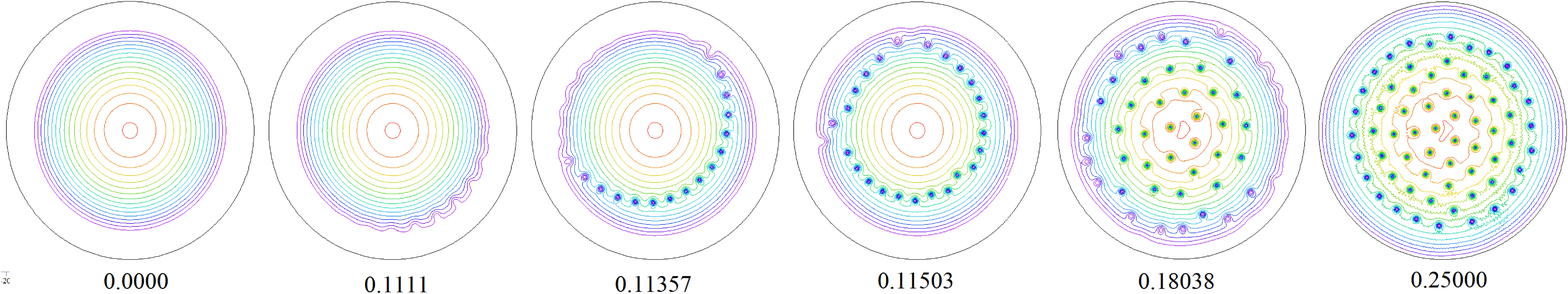}\\[0.25ex]
\includegraphics[width=0.97\textwidth]{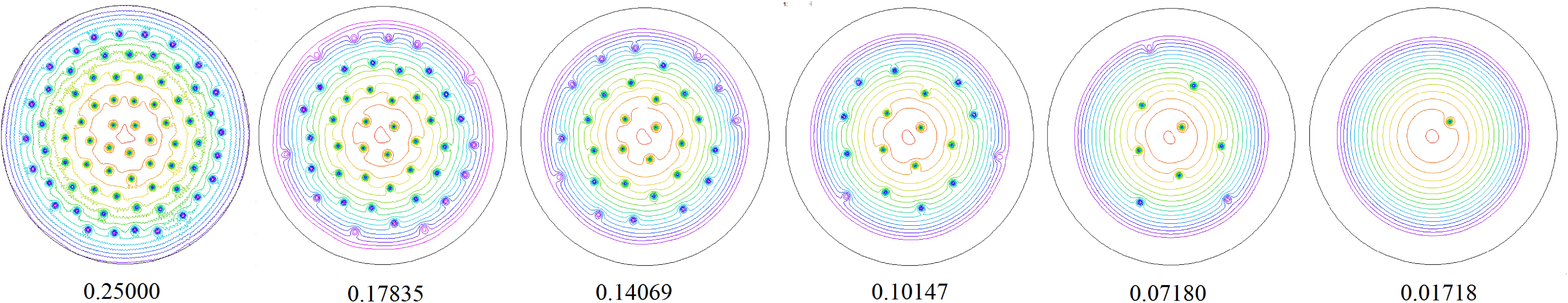}
\includegraphics[width=0.97\textwidth]{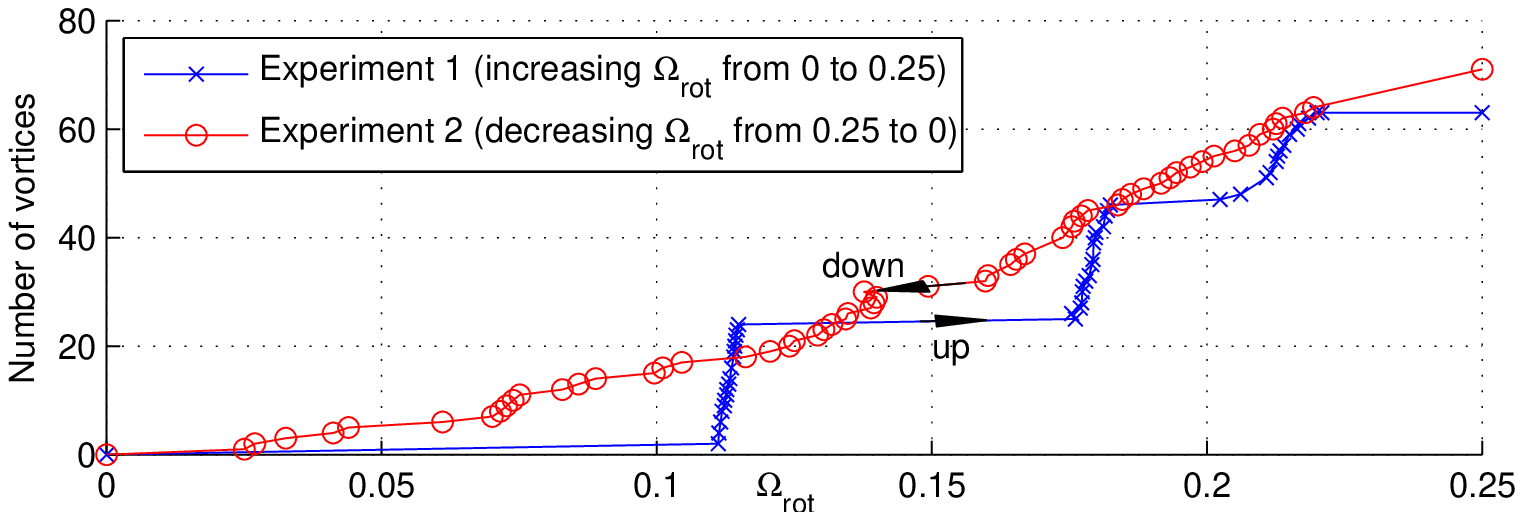}
\end{center}
\caption{(color online)
Top: snapshots of the simulation of (\ref{tk_eqn4}) with slowly varying
$\Omega_{\rm rot}$ (as indicated) and with
other parameters given by $\Omega_{\rm trap}=0.3538$ and $\mu=16.1$
(cf.~Ref.~\cite{dshall3}). Top row: $\Omega_{\rm rot}$ is slowly increased
from 0 to 0.25
according to the formula (\ref{Omega-up}).
Middle row: $\Omega_{\rm rot}$ is slowly decreased
from 0.25 to 0
according to the formula (\ref{Omega-down}).
Bottom row: number of vortices as a function of $\Omega_{\rm rot}$.
}
\label{fig:snapshots}
\end{figure}

For motivational purposes, we start our presentation with two direct
numerical experiments of Eq.~(\ref{tk_eqn4}).

\noindent
\textbf{Experiment 1: Slow Increase in Rotation.}
Let us allow for the rotation $\Omega_{\rm rot}$ to slowly
increase from 0 to 0.25.
More specifically, we start with $\Omega_{\rm rot}=0$, evolve the
system until $t=2000$ (so that it effectively reaches its ground state under
the imaginary time integration), then start increasing $\Omega_{\rm rot}$
linearly from 0 to 0.25 as $t$ varies from 2000 to 4000, and then
finally we keep $\Omega_{\rm rot}=0.25$ until $t=10000$. In full, this
experimental protocol can be summarized as:
\begin{equation}
\Omega_{\rm rot}\text{\texttt{\ = 0.25*min(max((t-2000)/2000,0),1)}}.
\label{Omega-up}
\end{equation}
We used {\tt FlexPDE} to simulate Eq.~(\ref{tk_eqn4}) with a uniform mesh
with around 13000 triangles and adaptive time stepping.

\noindent
\textbf{Experiment 2: Slow Decrease in Rotation and Hysteresis.}
Let us now slowly decrease $\Omega_{\rm rot}$ from 0.25 to 0. In
this complementary numerical experiment, we start with $\Omega_{\rm rot}=0.25$,
and dynamically evolve the system until $t=2000$ to allow it to relax to its
preferred vortex-lattice ground state profile. Then, we start decreasing
$\Omega_{\rm rot}$ linearly to 0 as $t$ varies from 2000 to 4000, then keep
$\Omega_{\rm rot}=0$ until $t=10000$. Mathematically again, this
procedure can be summarized as:
\begin{equation}
\Omega_{\rm rot}\text{\texttt{\ = 0.25*(1-min(max((t-2000)/2000,0),1))}}.
\label{Omega-down}
\end{equation}
Figure~\ref{fig:snapshots} presents a series of snapshots for each of the
described experiments. There are a number of interesting observations to
make here, as well as connections to provide with the discussion in the
previous section. As $\Omega_{\rm rot}$ is ramped up, up to a critical
rotational frequency no vortices arise. However, when they do arise (and
despite the weak ramping), a considerable number of vortices seems to emerge
asymmetrically (at first) yet nearly {\em simultaneously}. Gradually, as the rotation frequency increases, 
additional vorticity is ``elicited'' from the boundary, eventually
leading the configuration to self-organize into a triangular vortex lattice
as the final rotation frequency is reached. On the other hand, while the
frequency is decreased towards $\Omega_{\rm rot}=0$, we can see that
the process is clearly hysteretic, as for similar values of the rotation
frequency as in the top panel, a considerably larger number of vortices
appears to survive. This feature is most dramatic near the
$\Omega_{\rm rot}=0$ (e.g. in the next to last panel of the bottom row) where numerous
vortices appear to survive in this metastable dynamics, although clearly
this is far from the ground state for that rotation frequency. Although
there are numerous features that one may wish to explore on the basis of
this dynamical simulation and the
numerical results presented in the previous section,
the one that we will focus on below (following up on the discussion of the
earliest part of the previous section) concerns the instability dynamics
of the dissipative variant of the GPE for the
vortex-less steady state configuration. In particular, our expectation on the
basis of the above direct simulation, as well as from the stability results
presented is that a large $m$ mode is the one (predominantly) responsible
for the destabilization of the vortex-less state, 
as is indeed observed in Fig.~\ref{fig:snapshots}. We now
proceed to analyze this trait mathematically in more detail at the level
of Eq.~(\ref{tk_eqn4}).

\subsection{Asymptotic Analysis}

\begin{figure}[tb]
\begin{center}
\begin{tabular}{cc}
\includegraphics[height=5.0cm]{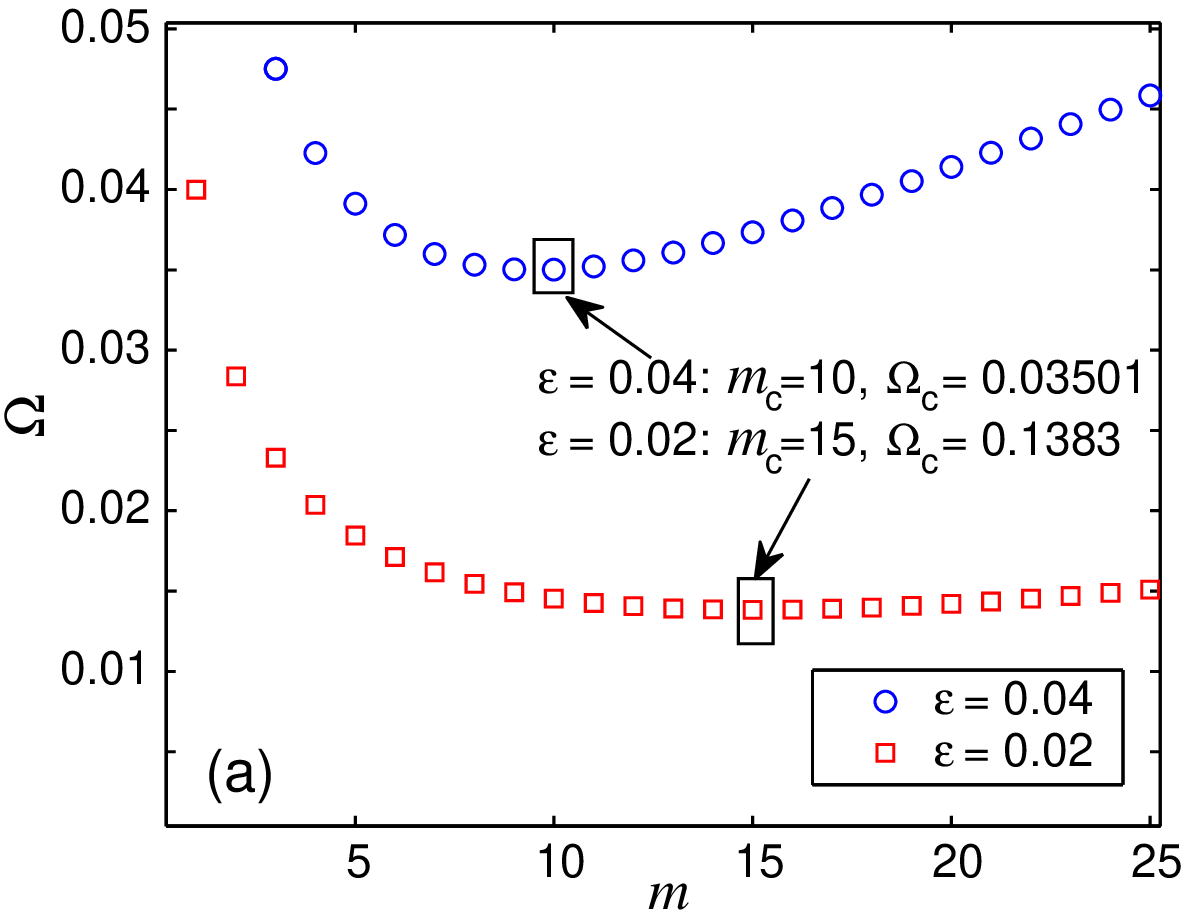}&
\includegraphics[height=5.0cm]{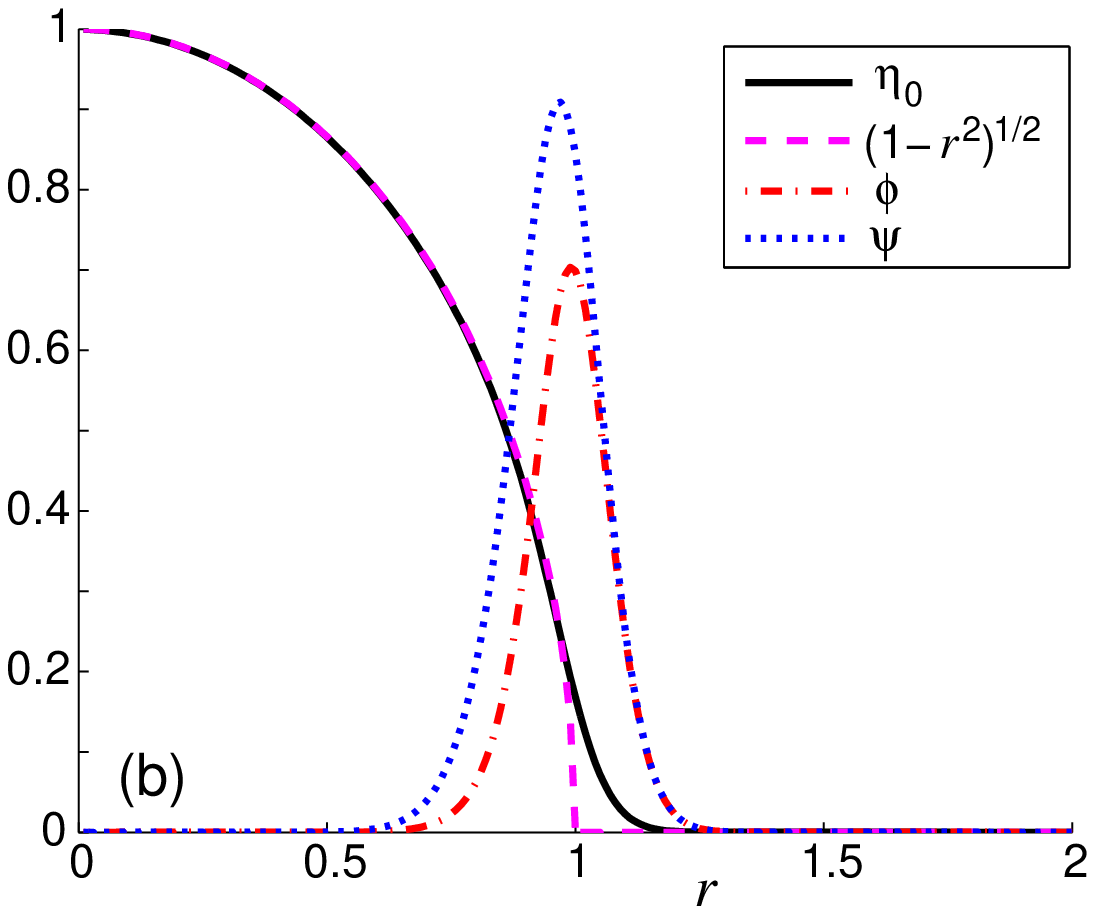}\\
\end{tabular}
\includegraphics[height=5.0cm]{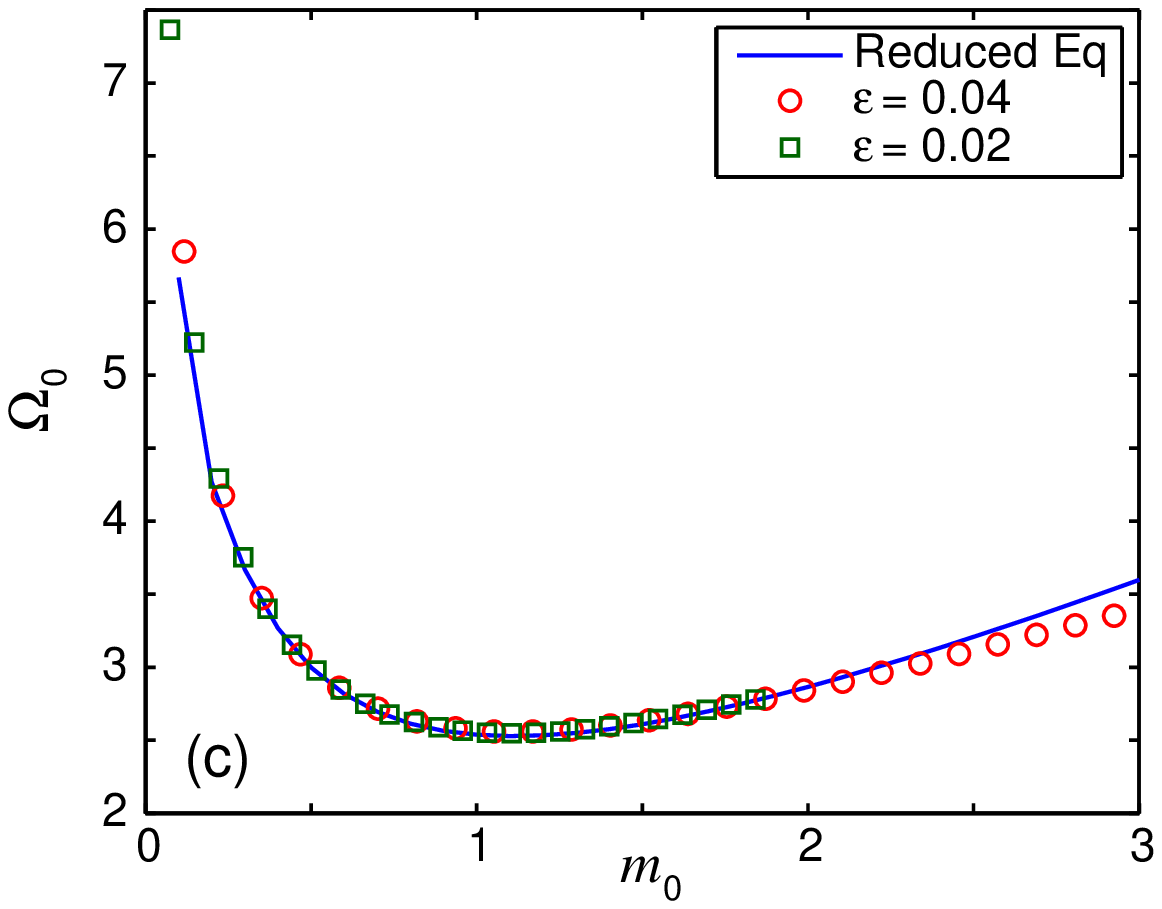}\\
\end{center}
\caption{(color online)
(a) Solution to Eq.~(\ref{origroast}) showing $\Omega$ as a function of $m$.
For a given mode $m$ on the horizontal axis, the vertical axis shows the threshold value of $\Omega$
for which this mode becomes unstable. The minimum of this graph is the overall threshold where the instability first sets
in as $\Omega$ is increased from zero.
(b) The profile of $\eta_0$, its Thomas-Fermi asymptotic approximation
$\eta_0 \sim \max((1-r^2),0)^{1/2}$,
as well as the profile of the eigenfunctions
corresponding to the problem (\ref{origroast}).
(c) Solution to the reduced system (\ref{reduced}) as compared with the full system (\ref{origroast})
}
\label{fig:omegam}
\end{figure}

To start our analysis, we rescale Eq.~(\ref{tk_eqn4}) as follows:
\begin{equation*}
x=\hat{x}\sqrt{\frac{\mu}{\frac{1}{2}\Omega_{\rm trap}^{2}}};\ \ u=\mu\hat {u};\
\ \ t=\frac{\hat{t}}{\mu}.\
\end{equation*}
After dropping the hats, we obtain
\begin{equation}
u_{t}=\varepsilon^{2}\Delta u+\left( 1-\left\vert x\right\vert ^{2}\right)
u-\left\vert u\right\vert ^{2}u-\Omega iu_{\theta},
\label{rescaled}
\end{equation}
where
\begin{equation}
\varepsilon=\frac{1}{2\mu}\sqrt{\Omega_{\rm trap}^{2}};
\text{ \ \ }
\Omega = \frac{\Omega_{\rm rot}}{\mu}.
\end{equation}
This is equivalent to the following real system of PDEs for the real and
imaginary parts of $u=v+iw$:
\begin{equation}
\left\{
\begin{array}{c}
v_{t}=\varepsilon^{2}\Delta v+\left( 1-\left\vert x\right\vert ^{2}\right)
v-v^{3}-w^{2}v+\Omega w_{\theta}, \\ [1.0ex]
w_{t}=\varepsilon^{2}\Delta w+\left( 1-\left\vert x\right\vert ^{2}\right)
w-w^{3}-v^{2}w-\Omega v_{\theta}.
\end{array}
\right.
\label{realsys}
\end{equation}
Let $u=\eta_{0}$ be the radially-symmetric vortex-less state which
satisfies:
\begin{equation}
0=\varepsilon^{2}\left( \eta_{0rr}+\frac{1}{r}\eta_{r}\right) +\left(
1-r^{2}\right) \eta_{0}-\eta_{0}^{3}.  \label{eta0}
\end{equation}
We linearize Eq.~(\ref{realsys})\ around $\eta_{0}$, so that
\begin{align*}
v & =\eta_{0}+e^{\lambda t}\phi(r), \\
w & =0+e^{\lambda t}\psi(r),
\end{align*}
to obtain the system
\begin{equation}
\left\{
\begin{array}{c}
\lambda\phi=\varepsilon^{2}\Delta\phi+\left( 1-r^{2}-3\eta_{0}^{2}\right)
\phi+\Omega\psi_{\theta}, \\ [2.0ex]
\lambda\psi=\varepsilon^{2}\Delta\psi+\left( 1-r^{2}-\eta_{0}^{2}\right)
\psi-\Omega\phi_{\theta}.
\end{array}
\right.
\end{equation}
We now use the radial-polar decomposition for the perturbations of the form:
\begin{equation*}
\phi(r,\theta)=e^{im\theta}\phi(r);\ \ \
\psi(r,\theta)=e^{im\theta}\psi\left( r\right),
\end{equation*}
to obtain
\begin{equation}
\left\{
\begin{array}{c}
\lambda\phi=\varepsilon^{2}\left( \phi_{rr}+\frac{1}{r}\phi_{r}-
\frac{m^{2}}{r^{2}}\phi\right) +\left( 1-r^{2}-3\eta_{0}^{2}\right)
\phi+im\Omega \psi, \\ [2.0ex]
\lambda\psi=\varepsilon^{2}\left( \psi_{rr}+\frac{1}{r}\psi_{r}-
\frac{m^{2}}{r^{2}}\psi\right) +\left( 1-r^{2}-\eta_{0}^{2}\right)
\psi-im\Omega\phi,
\end{array}
\right.
\end{equation}
Changing variables to 
$\psi=i\hat{\psi}$, 
and dropping the hat we then get a purely real system
\begin{equation}
\left\{
\begin{array}{c}
\lambda\phi=\varepsilon^{2}\left( \phi_{rr}+\frac{1}{r}\phi_{r}-\frac{m^{2}}
{r^{2}}\phi\right) +\left( 1-r^{2}-3\eta_{0}^{2}\right) \phi-m\Omega\psi,
\\ [2.0ex]
\lambda\psi=\varepsilon^{2}\left( \psi_{rr}+\frac{1}{r}\psi_{r}-\frac{m^{2}}
{r^{2}}\psi\right) +\left( 1-r^{2}-\eta_{0}^{2}\right) \psi-m\Omega\phi.
\end{array}
\right.
\end{equation}
It is known that $\lambda<0$ when $\Omega=0$ and $\lambda>0$ for
sufficiently large $\Omega$, as corroborated also by our numerical
computations in the previous section. We therefore seek the instability
threshold value for $\Omega$ for which $\lambda=0$. Thus, setting $\lambda=0$
in the eigenvalue problem, we obtain a modified eigenvalue problem for
$m\Omega$ of the form:
\begin{equation}
\left\{
\begin{array}{c}
m\Omega\psi=\varepsilon^{2}\left( \phi_{rr}+\frac{1}{r}\phi_{r}-\frac{m^{2}}
{r^{2}}\phi\right) +\left( 1-r^{2}-3\eta_{0}^{2}\right) \phi, \\ [2.0ex]
m\Omega\phi=\varepsilon^{2}\left( \psi_{rr}+\frac{1}{r}\psi_{r}-\frac{m^{2}}
{r^{2}}\psi\right) +\left( 1-r^{2}-\eta_{0}^{2}\right) \psi.
\end{array}
\right.  \label{origroast}
\end{equation}
To obtain an intuitive understanding of the situation ahead of the detailed
analysis, let us solve this problem numerically and then plot the graph of
$m$ vs.~$\Omega$. This is shown in Fig.~\ref{fig:omegam}(a) for
$\varepsilon=0.04$ and $\varepsilon=0.02$. Recall that the large
density/large chemical potential limit is associated with $\varepsilon
\rightarrow 0$, hence the choice of suitably small $\varepsilon$. We find
that this graph has a \emph{minimum} which corresponds to the smallest value
of $\Omega=\Omega_{c}$ for which the instability first manifests itself.
Critically, for our discussion, this minimum depends on $\varepsilon$. Our
goal is to characterize this minimum analytically, as well as to compute the
corresponding wave number $m=m_{c}$ which should approximate the
instability eigenmode that manifests itself as $\Omega$ increases and first
crosses past $\Omega_{c}$.

Figure~\ref{fig:omegam}(b)\ shows the plot of the leading eigenfunction for
$\varepsilon=0.02$ with $m=15$ (corresponding to the critical threshold
$\Omega_{c}=0.01383$). The relevant mode
 appears to be localized near $r\sim1$. To capture
this, we first have to resolve the corner layer of the steady state $\eta_0$
accurately near $r=1$. To that effect, we zoom in at $r=1$ and rescale
according to:
\begin{equation*}
r=1+\varepsilon^{2/3}y,\ \ \ \ \eta_{0}=\varepsilon^{1/3}U.
\end{equation*}
To leading order then we get from Eq.~(\ref{eta0})
\begin{equation}
U_{yy}=2yU+U^{3},
\label{poincare}
\end{equation}
which is a rescaled Painlev\'{e} II transcendent. It is
well-known~\cite{clarkson} that Eq.~(\ref{poincare}) admits a \emph{unique}
solution of the form
\begin{equation*}
U\sim\left\{
\begin{array}{c}
C\,{\rm Ai}\left( \sqrt{2}y\right) \text{ as }y\rightarrow\infty \\ [2.0ex]
\sqrt{-2y}\text{ \ \ as }y\rightarrow-\infty
\end{array}
\right.
\end{equation*}
where ${\rm Ai}$ is the Airy function and $C$ is a constant. 
Next, we use the same change of variables in the eigenvalue problem
(\ref{origroast}). We obtain then the reduced problem
\begin{equation}
\left\{
\begin{array}{c}
m_{0}\Omega_{0}\psi=\phi_{yy}-m_{0}^{2}\phi+\left( -2y-3U^{2}\right) \phi, \\ [2.0ex]
m_{0}\Omega_{0}\phi=\psi_{yy}-m_{0}^{2}\psi+\left( -2y-U^{2}\right) \psi,
\end{array}
\label{reduced}
\right.
\end{equation}
subject to the boundary conditions $\{\phi,\psi\}\rightarrow0$ as $\left\vert
y\right\vert \rightarrow\infty,$ where
\begin{equation}
m=\varepsilon^{-2/3}m_{0},\ \ \ \ \Omega=\varepsilon^{4/3}\Omega_{0}.
\label{scaling}
\end{equation}

The problem (\ref{reduced}) is solved numerically. The solution is shown in
Fig.~\ref{fig:omegam}(c). Superimposed are the solutions to the full
eigenvalue problem (\ref{origroast}) for $\varepsilon =0.02$ and
$\varepsilon =0.04$. It can thus be clearly observed that the scaling
(\ref{scaling}) is indeed correct.
From Fig.~\ref{fig:omegam}, the minimum is attained at around $\Omega
_{0,c}\approx 2.5$ (using the $\varepsilon =0.02$ curve). We now state this
result (a numerically assisted proof for the result in
provided in the appendix):

\textbf{Main Result.} \emph{There exist constants $\Omega_{0,c}$
and $m_{0,c}$ whose approximate values are}
\begin{equation*}
\Omega_{0,c}\approx 2.529,\ \ m_{0,c}\approx 1.111
\end{equation*}
\emph{such that the following is true. Let}
\begin{equation*}
\Omega_{c}=\varepsilon ^{4/3}\Omega_{0,c},\ \ \ m_{c}=\varepsilon
^{-2/3}m_{0,c}.
\end{equation*}
\emph{Then the vortex-less steady state (\ref{eta0}) of Eq.~(\ref{rescaled}) is
stable when $\Omega <\Omega_{c}$ and becomes unstable as $\Omega$
crosses $\Omega_{c}$. The fastest-growing unstable mode
corresponds to the oscillations of the boundary with the mode $m_{c}$.}

Note that in terms of the original variables of Eq.~(\ref{tk_eqn4}), the
critical threshold is given by
\begin{align}
\Omega_{{\rm rot},c} &
=\Omega_{0,c}2^{-4/3}\mu^{-1/3}\Omega_{\rm trap}^{4/3}\approx1.0036\mu^{-1/3}
\Omega_{\rm trap}^{4/3}; \\
m_{c} & =m_{0,c}2^{2/3}\mu^{2/3}\Omega_{\rm trap}^{-2/3}\approx1.76\mu
^{2/3}\Omega_{\rm trap}^{-2/3}.
\end{align}

In Experiment 1 of the previous section depicted in Fig.~\ref{fig:snapshots},
we had $\mu=16.1$ and $\Omega_{\rm trap}=0.3538$;
these numbers correspond to the experimentally
relevant setting of the work of Ref.~\cite{dshall3}. This yields
$\Omega_{{\rm rot},c}\approx0.10$ and $m_{c}\approx22.11$.
This is in excellent agreement
with the actual numerical simulations. 
In the top row of the figure, the instability becomes apparent around
$\Omega_{\rm rot}\approx0.11$. The instability actually sets in shortly
prior to this, but it takes some time for it to fully mature. Once the
instability is fully developed (fourth snapshot from the left), it results
in 24 vortices, in fair agreement with the predicted value of
$m_{c}\approx22$ (it should be kept in mind that some of these 
vortices may be in the periphery of the cloud and hence may not
be discernible in the density profile shown).


It is interesting finally to connect the results, e.g., with those of
Figs.~\ref{tkfig2}--\ref{tkfig4}.
In particular, as indicated in these results the left
panel, e.g., of Fig.~\ref{tkfig2} corresponds to $\mu=5$, while the right one
to $\mu=10$. According to the above scaling in the former case
$\Omega_{{\rm rot},c} \approx 0.343$, while in the latter case
$\Omega_{{\rm rot},c} \approx 0.272$.
It can be seen that these critical points are in
excellent agreement with the crossing points (from positive to negative
energy modes) of the Hamiltonian system in Fig.~\ref{tkfig2}, which involves
the case of $\gamma=0$. They are also in excellent agreement with the
stability thresholds of the dissipative system used in
Figs.~\ref{tkfig3} and \ref{tkfig4}, although the latter only use $\gamma=0.01$.
This comparison is given to also, a posteriori, justify the use of the
analysis in the framework of the overdamped
limit of Eq.~(\ref{tk_eqn4}) in the present context.

\section{Conclusions and Future Challenges}
\label{sec:conclu}

In summary, in the present work, we have explored the instability in the
presence of rotation of a dissipative variant of the GPE. We have connected
this instability to the emergence of negative energy (or Krein signature)
modes and the phenomenon of energetic (but not dynamical) instability of the
radial vortex-less profile in the corresponding Hamiltonian system in the
absence of dissipation. We have also connected it to the emergence of a real
eigenvalue corresponding to suitably large $m$ (i.e., azimuthal order) for
the dissipative system. Moreover, we have argued that this $m$ should be
independent of $\gamma$, but should depend on the trapping frequency and on
the chemical potential (i.e., the maximal atomic density) of the system. We have
systematically developed a scaling law that provides the critical rotation
frequency as a function of these parameters. We have confirmed the
connection of this scaling with (a) the imaginary time (``overdamped'')
model used; (b) the
original Hamiltonian model and (c) the intermediate between the two
dissipative Gross-Pitaevskii equation (DGPE) model.
Using direct numerical simulations we have corroborated the prediction
of our asymptotic analysis and observed that these unstable modes
with high mode number $m$ indeed nucleate a large number of vortices at the
periphery of the atomic cloud.
However, we have found that despite the large number of vortices at
the periphery, for the small values of $\gamma$ chiefly considered
herein in the DGPE (yet somewhat larger than the
ones that have been claimed as relevant for realistic 
experimental settings; see for a recent discussion~\cite{dongyan})
few isolated vortices are pulled in sequentially towards
the center of the cloud. The process whereby single vortices are
singled out from the multi-vortex ``necklace'' at the periphery is a
pattern selection mechanism based on symmetry-breaking.
This sequential recruiting of peripheral vortices towards the cloud 
center saturates when reaching a highly symmetric configuration with
a few vortices at the center of the cloud that is {\em stable} for the chosen
parameters (chemical potential and rotation rate normalized by trap
strength). 

These results open a number of interesting directions for further exploration. 
Admittedly, our approach to the vortex nucleation problem is rather 
different from that of earlier works; see e.g.~\cite{lundh2} and
the relevant discussion of Ref.~\cite{becbook2}. Nevertheless, it would
be particularly interesting to explore in current experimental
settings (such as e.g.~\cite{dshall3}) whether a value of 
$\gamma$ can be ``inferred'' (e.g. from the spiraling motion of a
vortex; see e.g. the relevant discussion of~\cite{dongyan}).
Based on such a value, our analysis and computation could provide
diagnostics both for which eigenmode will cause the instability of
the vortex-less state and for which asymptotic state may be 
experimentally observed. 

Our observations also raise a related problem. Clearly, for different
values of $\gamma$ ranging from the underdamped limit
of Figs.~\ref{fig:snap_film_mu05}--\ref{fig:snap_film_mu05_Nv5}
to the overdamped one of Fig.~\ref{fig:snapshots},
(for same trap strength and chemical potential) we have
conclusively argued that the same eigenmode and the same critical
frequency are generically responsible for the observed
instability. Yet, the 
instability manifestation has dramatically different outcomes
for the different limits. This naturally prompts the question:
what is the favored asymptotic state (depending on the value
of $\gamma$) and how do we get there? The first and perhaps
simpler question (that has been previously considered; see e.g.~\cite{castin}
for an early example) is
presumably one of energetic comparisons of the states containing
different numbers of vortices (incorporating their angular momentum
contributions). The second and, arguably, more 
difficult question is one of transition state pathways that
enable the nucleation of different multi-vortex configurations.
The latter may be particularly worth exploring, especially since our results
indicate that they will be dependent on features such as the
thermal coupling parameter $\gamma$.

\Appendix
\section{A Numerically Assisted Proof Of The Main Result}

Define the operators
\begin{equation*}
\left\{
\begin{array}{rl}
L_{1}(\phi)\hskip-0.25cm&:=\phi_{yy}+\left( -2y-3U^{2}\right) \phi, \\[2.0ex]
L_{2}(\phi)\hskip-0.25cm&:=\phi_{yy}+\left( -2y-U^{2}\right) \phi.
\end{array}
\right.
\end{equation*}
As the first step, we show that both $L_{1}$ and $L_{2}$ are negative
operators. First, note that that both are self-adjoint so it suffices to
show that all eigenvalues are real non-positive. To show that $L_{2}$ is
negative, simply note that $L_{2}(U)=0$. Since $U$ is positive, Sturm's
eigenvalue oscillation theorem then implies that $\lambda=0$ is the largest
eigenvalue of $L_{2}$, hence $L_{2}$ is negative. The negativity of $L_{1}$
follows from the fact that $2y+3U^{2}>0$ for all $y$;
see Ref.~\cite{pelinovskios}.

The problem (\ref{reduced})\ may then be reformulated as
\begin{equation}
m_{0}\Omega_{0}=\pm\sqrt{\mu};\ \ \
\mu\phi=(L_{1}-m_{0}^{2})\cdot(L_{2}-m_{0}^{2})\,\phi.  \label{mu}
\end{equation}
It follows by the negativity of $L_{1}-m_{0}^{2}$ and $L_{2}-m_{0}^{2}$ that
$\mu$ is real and positive so that the curve $\Omega_{0}=\Omega_{0}(m_{0})$
is well defined.

Finally, we show that the curve $\Omega_{0}=\Omega_{0}(m_{0})$ has a minimum
for some strictly positive value of $m_{0}$. For large $m_{0}$ we find that
$\mu\sim m_{0}^{4}$ and hence $\Omega_{0}\sim m_{0}$. On the other hand,
numerical computations of (\ref{mu})\ with $m_{0}=0$ yield $\mu=0.1576654>0$.
It follows that $\Omega_{0}\sim0.39707/m_{0}$ as $m_{0}\rightarrow0^{+}$.
Thus $\Omega_{0}$ blows up at the endpoints $m_{0}\rightarrow0^{+}$ and
$m_{0}\rightarrow\infty$, which shows that this curve indeed has a minimum.

\bigskip
\textbf{Acknowledgments.}
We are grateful to Dmitry Pelinovsky for useful discussions and
for insights leading to the proof of the Main Result (Appendix A).
R.C.G.~acknowledges support from DMS-1309035.
P.G.K.~acknowledges support from the National Science Foundation under
grants  DMS-1312856, from ERC and FP7-People under grant IRSES-605096,
from the US-AFOSR under grant FA9550-12-10332, and from the Binational
(US-Israel) Science Foundation through grant 2010239.
P.G.K.'s work at Los Alamos is supported in part
by the U.S. Department of Energy.
T.K.~was supported by NSERC Discovery Grant No.~RGPIN-33798 and
Accelerator Supplement Grant No.~RGPAS/461907.
%


\end{document}